\documentclass[preprint,12pt]{elsarticle}
\usepackage[a4paper, left=2cm, right=2cm, top=2.5cm, bottom=2.5cm]{geometry}

\usepackage{graphicx}
\usepackage{dcolumn}
\usepackage{bm}
\usepackage{amssymb}
\usepackage{amsmath}
\usepackage{color}
\usepackage{physics}
\biboptions{sort&compress}
\usepackage{soul}
\usepackage{hyperref}
\usepackage{xcolor}
\usepackage{cancel}

\hypersetup{
    colorlinks=true, linkcolor=blue, citecolor=blue,
    filecolor=blue, urlcolor=blue, breaklinks=true
}

\newcommand{\orcid}[1]{\href{https://orcid.org/#1}
  {\includegraphics[width=7pt]{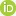}}}

\journal{arXiv}

\begin{document}

\begin{frontmatter}
\title{Non-Markovian Light–Matter Dynamics in the Time Fractional Jaynes–Cummings Model with Modulated Coupling} 

\author{Enrique Chipicoski Gabrick$^{1*\orcid{0000-0002-8407-7675}}$}
\author{Thiago Takaji Tsutsui$^{2,**\orcid{0009-0001-1654-0330}}$}
\author{Danilo Cius$^{1\orcid{0000-0002-4177-1237}}$}
\author{Ervin Kaminski Lenzi$^{3\orcid{0000-0003-3853-1790}}$}
\author{Antonio S. M. de Castro$^{2,4,\orcid{0000-0002-1521-9342}}$}
\author{Fabiano M. Andrade$^{2,5,6\orcid{0000-0001-5383-6168}}$}

\address{
$^1$Institute of Physics, University of São Paulo, 
05508-090 São Paulo, São Paulo, Brazil.\\
$^2$Graduate Program in Science, State University of Ponta Grossa,
84030-900, Ponta Grossa, Paraná, Brazil.\\
$^3$Department of Physics, State University of Maring\'a, 87020-900, Maring\'a, Paraná, Brazil.\\
$^4$Department of Physics, State University of Ponta Grossa
  84030-900, Ponta Grossa, Paraná, Brazil.\\
$^5$ Department of Mathematics and Statistics, State University of Ponta Grossa
  84030-900, Ponta Grossa, Paraná, Brazil.\\
$^6$Departament of Physics,
  Federal University of Paran\'a,
  81531-980 Curitiba, Paraná, Brazil.}

\cortext[cor]{ecgabrick@gmail.com}
\cortext[cor]{takajitsutsui@gmail.com}

\date{\today}

\begin{abstract}
We investigate the fractional time description of a generalized quantum light-matter system modeled by a time-dependent Jaynes-Cummings (JC) interaction, with different coupling types: constant, linear, exponential, and sinusoidal. 
Two formulations of the time fractional Schr\"odinger equation (TFSE) are examined, with a focus on their impact on population inversion and entanglement. 
Our findings highlight that the introduction of fractional order introduces memory effects, associated with damped oscillations and asymptotic decay. 
Furthermore, we find that the time-dependent couplings, combined with distinct fractional formulations, influence how these effects occur, ultimately resulting in high or low entanglement. 
A key finding of our work is that, under sinusoidal coupling, non-periodic dynamics is preserved for both formulations of the TFSE; however, within a certain range, the fractional order can act as a control mechanism for the non-periodic evolution.\\

\noindent doi: \href{https://doi.org/10.1016/j.cnsns.2026.109912}
{10.1016/j.cnsns.2026.109912}
\end{abstract}


\end{frontmatter}

\sloppy

\section{Introduction}
Fractional calculus (FC) has been a powerful alternative framework for studying various physical scenarios \cite{TenreiroAdvances}. 
It has emerged as an extension of integer-order derivatives to non-integer-order derivatives \cite{TenreiroHistory}. These extensions incorporate many new results with non-trivial consequences, including memory effects and long-range correlations \cite{ali2020}, which are connected to non-Markovian processes \cite{Evangelista2018}. 
FC has been applied in various fields, e.g., the study of infectious diseases \cite{Gabrick2023inf,TenreiroHIV}, 
complex viscoelastic phenomena  \cite{pandey2016physical,bagley1983theoretical}, 
wave propagation in porous media \cite{chen2016causal,cai2018survey},
electrical spectroscopy impedance \cite{rosseto2022frequency}, gas transport in heterogeneous environments \cite{chang2018time,chang2019spatial,pandey2016connecting},
{
to name a few.}

Among its recent advances, fractional calculus (FC) has been integrated into quantum mechanics by replacing integer derivatives in the Schrödinger equation with non-integer-order operators. 
This extension can be traced back to the pioneering works of Laskin \cite{laskin2000,PhysRevE.66.056108} and Naber \cite{Naber2004}. 
In Naber's contribution, the time fractional Schrödinger equation (TFSE) is formulated in two distinct ways: the first approach raises the imaginary unit to the same power as the fractional derivative, while the second approach retains the imaginary term in its original form.  
In both frameworks, the evolution is non-unitary \cite{Cius2022}, although the specific mechanisms through which this occurs differ. 
The first formulation is linked to a Wick rotation, resulting in {
oscillatory effects}. 
In contrast, the second formulation is associated with dissipation \cite{Wei2024}.

In the context of fractional time operators, Naber employed the Caputo fractional derivative in the TFSE formulation \cite{Naber2004}, making the TFSE suitable for describing non-Markovian processes \cite{Rivas2010, Zou2015} characterized by a power-law memory kernel.  
More specifically, FC has been applied to study different scenarios in quantum mechanics \cite{Ma2025}, 
e.g., 
time-dependent quantum potentials \cite{Gabrick2023frac},  
anomalous diffusion in a three-level system \cite{lenzi2022schrodinger}, 
Lévy flights over quantum paths \cite{Laskin2007}, 
and quantum comb structures \cite{Iomin2011}. 
In the case of quantum optics, 
the FC was explored both theoretically \cite{Longhi2015} and experimentally \cite{Huang2005}.

In this framework, the interaction of a two-level system with a quantized electromagnetic field is modeled with a Jaynes-Cummings (JC) interaction \cite{JAYNES1963,LARSON2021}. 
This model has garnered attention in the fractional quantum mechanics community, having been studied in diverse contexts and with various objectives.
Equipped with FC, the study of a two-level system interacting with a semi-classical light field was conducted by Lu and Yu \cite{lu2017,Lu2018}. 
Differential equations similar to the
ones in the JC model arise, although the physical interpretation differs.
Subsequently, El Anouz {et al.} \cite{ElAnouz2020} analyzed the Fisher information of the JC model within the FC framework.
Zu {et al.} \cite{Zu2021} studied single and double JC,
with a focus on memory effects and entanglement.
In 2022, Zu and {Yu} \cite{zu2022} revisited the problem using a limit-based approach,
resulting in a probability-conserving evolution for the system.
El-Hadidy {et al.} \cite{El-Hadidy2023} examined the communication efficiency of the Anti-JC model.
In more recent developments, a time fractional JC model was investigated in Ref. \cite{Cius2024}, where a specific protocol was considered to obtain the unitary time-evolution operator when the Hamiltonian operator is traceless, 
based on a non-Hermitian map \cite{Cius2022}.

In this work, we study a fractional generalization of this atom-field interaction, where the light-matter coupling changes with time, 
modeling it with the time-dependent JC (TDJC) model \cite{Schlicher1989,Prants1992,Joshi1993,Larson2003,Larson2003a}.
The TDJC introduces changes in the standard JC model, 
such as the fastening of the Rabi oscillations (RO) \cite{Joshi1993}. 
Our primary goal is to study how these modifications occur in the FC context. 
With this, we can see how the effects associated with the FC permeate the physical scenarios -- such as atomic motion, spatial variations in the field, and transient effects in the cavity -- that are modeled via time-dependent coupling.
To achieve this objective, we calculate the atomic population inversion and atom-field entanglement, as measured by the von Neumann entropy (VNE).
{Beyond its mathematical formulation, the TFSE provides an appropriate framework for describing memory effects in both formulations, $i^\alpha$ \cite{Zu2021}, for $\alpha \neq 1$, and $i^{1}$ \cite{Zu2025}. 
The latter, in particular, is beneficial for modeling the non-Markovian evolution of dissipative systems \cite{Breuer2016}.
Our approach to the JC is similar to recent work that frames the problem from this perspective, as seen in \cite{Wei2023,Wei2024,Zu2021,Zu2025}, however, we extended it by considering different couplings, both TFSE formulations, and obtaining new results concerning entanglement.}
{We emphasize that our goal is not to quantify non-Markovianity, but to study how FC-induced memory effects impact the dynamics in the TDJC model.
A comprehensive treatment of non-Markovianity is given in Ref. \cite{Breuer2016}, whereas other approaches to memory effects in the JC model are addressed from other perspectives in Refs. \cite{Madsen2011,Li2010,Monteiro2025}.
}

Given the applications of the JC model in quantum computing {\cite{Monroe1995,Azuma2011}}, these new perspectives may contribute to its contextualization in different perspectives.
For example, the sinusoidal coupling in the TDJC model describes an atom traversing a standing-wave cavity mode \cite{Schlicher1989,LARSON2021}.
In this scenario, the TFSE can incorporate memory effects into the physical picture, possibly facilitating an experimental verification of FC in cavity quantum electrodynamics (QED).

Quantifying atomic probabilities and entanglement poses a problem in straightforward approaches to quantum mechanics utilizing FC, since the fractional time derivative leads to the 
non-conservation of probability \cite{Laskin2017}. 
There are two ways to deal with this paradigm: 
(a) study the effects of FC  with a focus on the mathematical framework 
\cite{lu2017,Lu2018,Gabrick2023frac}{,} while acknowledging that physical interpretations remain largely heuristic,
or (b) find ways to reconcile the theory with standard quantum mechanics,
by proposing alternatives to the usual Caputo derivative \cite{zu2022}, with a unitary description \cite{Cius2022,Cius2024}, or normalizing the density operator \cite{El-Hadidy2023,ElAnouz2020,Zu2021} to ensure a well-defined statistical interpretation. 
Our approach is mainly the second,
applying a normalization to the observables and density matrix \cite{SERGI2013}.
References \cite{zu2022} and \cite{Cius2024}, employed alternative approaches to the same problem,
but their technique leads to different results compared to the normalization. 
During the development of our research, we became aware of the work published by Zu and Yu \cite{Zu2025}. 
However, our results differ from theirs, since we study different coupling configurations in the JC model and also consider the off-resonance condition. 
Remarkably, our investigation of the power of the imaginary unit yields outcomes that agree with those derived by Zu and Yu, although within a different framework.

We investigate four distinct couplings for JC dynamics: constant, linear, exponential, and sinusoidal. 
For the constant coupling case, we obtain analytical solutions that demonstrate the non-conservation of probability and highlight the role of the detuning parameter in this behavior. 
When the Wick rotation is employed, we observe oscillations with decreasing amplitude around a fixed value in the quantities considered.
On the other hand,  when the standard form of the imaginary unit is preserved, the oscillations eventually cease after a specific time for all the couplings, 
decaying to an asymptotic value. 
These effects are influenced both quantitatively and qualitatively by time-dependent couplings, ultimately resulting in high or low entanglement.  
Notably, in the sinusoidal form, our results suggest a non-periodic motion in the measured physical quantities, regardless of the specific representation of the imaginary unity. 
{ Furthermore, when the fractional order is reduced below $0.7$, the non-periodic behavior transitions into a dynamics with a smoother aperiodicity.}

We organize the paper as follows.
In Sec. \ref{sec:jc}, we present the JC and its time-dependent generalization.
The population inversion and the VNE are introduced. 
Subsequently, in Sec. \ref{sec:frac}, the FC formalism is established. 
The time fractional TDJC is explored in Sec. \ref{sec:frac_jc}, for constant, linear, exponential, and sinusoidal modulations of the coupling. 
The influence of the {system's} parameters on the {normalization parameter} is briefly discussed {for the constant coupling case}.
Finally, we present our conclusions in Sec. \ref{sec:conc}.

\section{Time-dependent Jaynes-Cummings model}
\label{sec:jc}
The JC model is a cornerstone in quantum optics, 
describing the interaction of a two-level atom with a quantized field mode under the rotating-wave approximation \cite{JAYNES1963}. 
The model is associated with different phenomena regarding the light-matter interaction,
such as 
atom-field entanglement \cite{Aravind1984,Phoenix1988,Phoenix1991}, 
RO \cite{Cummings1965,Eberly1980} and their
 collapses and revivals \cite{Cummings1965,Eberly1980}, 
 {which have been}
experimentally verified in the context of cavity QED \cite{MESCHEDE1985,REMPE1987,Brune1996}.  
For further information, we refer to the review conducted by Larson and Mavrogordatos \cite{LARSON2021} 
and the references therein.
The TDJC model extends the usual framework by allowing parameters of the system,
otherwise constant, to vary with time \cite{DeCastro2023,Larson2003,Larson2003a}.  
This enhances the scope of the system, eventually encompassing new physical scenarios,
such as atomic motion \cite{Schlicher1989,Fang1998}, transient effects \cite{Prants1992,Dasgupta1999}, and varying field intensities \cite{Joshi1993}.
According to Ref. \cite{Maldonado-Mundo2012}, a time-dependent coupling situation could be realized in cavity QED, with variations in the atom's position.

A priori, the atom-field coupling parameter ($\lambda$) {could} be considered constant. 
However, in this work, we focus on a broader scenario, i.e., $\lambda=\lambda(t)$,   
considering on- and off-resonance contexts. 
In this sense, the TDJC Hamiltonian is given by 
\begin{equation}
\label{eq:JC_Hamiltonian_full}
    \hat{H} (t) = \frac{1}{2} \omega \hat{\sigma}_z + \nu \hat{a}^\dagger \hat{a} 
    +   \lambda(t) (\hat{\sigma}_{+}\hat{a}+\hat{\sigma}_{-}\hat{a}^{\dagger}),
\end{equation}
where we assume $\hbar =1$, without loss of generality.
This extension not only leads to significant changes in the system’s dynamics but can also be applied from a control perspective in the JC model with fractional time, as suggested in Ref. \cite{Wei2024}.
We use $\omega$ and $\nu$ to represent the atomic transition frequency and the cavity mode frequency, respectively.
The creation ($\hat{a}^\dagger$) and annihilation operators ($\hat{a}$) 
act on the cavity states in the Fock basis $\left\{|n\rangle \right\}$. 
The atom is effectively treated as a two-level system with $|e\rangle$ ($|g\rangle$) representing the excited (ground) state, $\sigma_{+}=\ketbra{e}{g}$ and $\sigma_{-}=\ketbra{g}{e}$ are the raising and lowering operators, and $\sigma_{z}=\ketbra{e}{e}-\ketbra{g}{g}$ is the atomic inversion operator.

In what follows, we consider the initial state as
\begin{equation} \label{eq:init_state}
|\Psi(0)\rangle =a(0)|e,n\rangle+b(0) |g,n+1\rangle,
\end{equation}
which implies an evolved state in the form
\begin{equation} \label{eq:fock_state}
    |\Psi(t)\rangle =a(t)|e,n\rangle+b(t)|g,n+1\rangle.
\end{equation}
{In standard quantum mechanics, a closed quantum system evolves unitarily in time, and consequently, the total probability remains equal to one
for a normalized state.}

The system {under scrutiny} can be described by the effective time-dependent Hamiltonian $\hat{V}(t)$ {in the form}
\begin{equation} \label{eq:int_hamiltonian_jc}
    \hat{V} (t) =  \frac{1}{2} \Delta \hat{\sigma}_z +
    \lambda(t) (\hat{\sigma}_{+}\hat{a}+\hat{\sigma}_{-}\hat{a}^{\dagger}),
\end{equation}
with $\Delta=\omega-\nu$ as the detuning. 
The amplitudes $a(t)$ and $b(t)$ are obtained from the solution of the Schr\"odinger equation in the form 
\begin{equation}
    i \frac{d}{dt} |\Psi(t)\rangle = \hat{V} (t)  |\Psi(t)\rangle.
    \label{standard_schrodinger}
\end{equation} 
Eqs. {\eqref{eq:fock_state}, } \eqref{eq:int_hamiltonian_jc} and \eqref{standard_schrodinger}, 
{lead to} the system of linear differential
equations with time-dependent coefficients: 
\begin{equation}
    \begin{aligned}
        i  \frac{d}{d t} a(t)&=&\lambda(t)b(t)\sqrt{n+1}+\frac{1}{2}\Delta a(t), \label{eqs_standard} \\ 
        i  \frac{d}{d t} b(t)&=&\lambda(t)a(t)\sqrt{n+1}-\frac{1}{2}\Delta b(t).
    \end{aligned}
\end{equation}
Therefore, the evolution of $a(t)$ and $b(t)$ is completely determined by the solution of the system, Eq. \eqref{eqs_standard}.  
These quantities allow us to determine the atomic population inversion 
\begin{equation}
    W(t)=|a(t)|^2-|b(t)|^2. 
\end{equation}
The population inversion, which {is} experimentally measurable \cite{REMPE1987}, {serves as a standard subject of examination} when accounting for the JC model and its extensions \cite{Arroyo-Correa1990}.

Moreover, we also study the VNE for the atomic subsystem \cite{VonNeumann1927b}, 
described by the reduced density operator
\begin{equation}
    \hat{\rho}_A (t)=\tr_F [\hat{\rho}(t)] = \sum_{n=0}^{\infty}\langle n|\hat{\rho}(t) |n\rangle.
\end{equation}
By definition, $\hat{\rho}(t)=\ketbra{\Psi(t)}$ represents the density operator of the joint atom-cavity system.
For a bipartite quantum system and pure global state, the VNE is a good measure of entanglement \cite{Nielsen2010}, 
a quantum correlation in the atom-cavity system associated with the non-separability of the quantum state  \cite{Horodecki2009}.
The VNE is given by
\begin{equation} \label{eq:vne}  
     S^{A}(t) = -\sum_i \mu_i(t) \log_2{\mu_i(t)}, 
\end{equation}   
where $\mu_i(t)$ are the eigenvalues of $\hat{\rho}_A(t)$.
For the initial state in the form of Eq. \eqref{eq:init_state}, 
we can further write it as 
\begin{equation} \label{eq:vne_fock}  
     S^{A}(t) = - |a(t)|^2 \log_2{|a(t)|^2}-  |b(t)|^2 \log_2{|b(t)|^2}.
\end{equation}   

{For an arbitrary time-dependent coupling $\lambda(t)$,  closed-form analytical solutions for $a(t)$ and $b(t)$ are generally unavailable. 
Nevertheless, certain functional forms of $\lambda(t)$ do admit exact solutions \cite{Prants1992,Dasgupta1999}.}
Assuming $\lambda(t) = \lambda_0$ (constant coupling), Eq. \eqref{eqs_standard} can be decoupled and easily solved. 
For an arbitrary initial value problem with $a(0) \equiv a_0$ and $b(0) \equiv b_0$, their corresponding solutions take the following forms:
{
For an arbitrary initial value problem with} $a(0) \equiv a_0$ and $b(0) \equiv b_0$, {their corresponding solutions} take the following forms:
    \begin{equation}
    \begin{aligned}
    a(t) = 
    \frac{a_0}{2} \left(e^{-i \Omega_n t}+ e^{i \Omega_n t}\right)
     +  \frac{\left[a_0 \left(\frac{\Delta}{2}\right)+b_0\lambda_0 \sqrt{n+1}\right]}{2 \Omega_n}  
    \left(e^{-i \Omega_n t}- e^{i \Omega_n t}\right), \\
     b(t)  = 
     \frac{b_0}{2} \left(e^{-i\Omega_n t}+ e^{i\Omega_n t}\right) 
     + \frac{\left[a_0\lambda_0 \sqrt{n+1} -b_0 \left(\frac{\Delta}{2}\right)\right]}{2 \Omega_n} 
     \left(e^{-i \Omega_n t} - e^{i \Omega_n t}\right), \label{solucao_inteiro}
    \end{aligned}
    \end{equation}
where $\Omega_n=\sqrt{\Delta^2/4+\lambda_0^2(n+1)}$. 
For this case, it is straightforward that {$|a(t)|^2 + |b(t)|^2 = 1$}, in accordance with the requirement of unitary
evolution for the quantum state $|\Psi(t)\rangle$. 

\section{Fractional time calculus} 
\label{sec:frac}

The analysis of fractional dynamics in the context of the TFSE can be carried out using two distinct
approaches \cite{Naber2004}. 
The first is to raise $i$ to the same order as the fractional operator ($\alpha$), i.e., $i^\alpha$, through the action of a Wick rotation. 
{The main effect of this approach is the appearance of oscillations in quantum dynamics.}
On the other hand, the TFSE can be written using only $i$, {
including dissipative effects.} 
Both approaches can provide novel aspects in the TDJC model dynamics and can be simultaneously incorporated by writing the TFSE in the form
{
\begin{equation} \label{eq:ftse}
    i^\beta \frac{\partial^\alpha}{\partial t^\alpha} |\Psi_{\alpha,\beta}(t)\rangle=\hat{V}(t)|\Psi_{\alpha,\beta}(t)\rangle.
\end{equation}
}
This modification enriches the fractional solutions by introducing a new control parameter that accounts for the different effects arising from both definitions of the TFSE \cite{Cius2022}. 
{
Cases where $\beta \neq \alpha$ and $\beta \neq 1$ simultaneously combine oscillatory ($\beta = \alpha$) and dissipative ($\beta = 1$) dynamics. This regime represents a novel area for TFSE simulations; however, a detailed discussion is lengthy and will be addressed in future work.}

{
Henceforth, we shall denote 
\begin{equation} \label{eq:norm}
 \mathcal{P}_{\alpha,\beta}(t) = \langle\Psi_{\alpha,\beta}(t)|\Psi_{\alpha,\beta}(t)\rangle ,   
\end{equation}
which \textit{cannot be interpreted as the standard total probability}, except when $\alpha=\beta=1$, a point that will be discussed in the next section.
Nonetheless, $\mathcal{P}_{\alpha,\beta}(t)$ serves as a normalization parameter for both population inversion and VNE.
}
As a fractional operator, we employ the Caputo fractional time derivative \cite{Evangelista2018}{. For $0<\alpha<1$, the derivative is defined as} 
{\begin{equation} \label{eq:caputo}
    \frac{\partial^\alpha}{\partial t^\alpha} |\Psi_{\alpha,\beta}(t)\rangle=\frac{1}{\Gamma(1-\alpha)} \int_0^t d t^{\prime} \frac{1}{\left(t-t^{\prime}\right)^\alpha} \frac{\partial}{\partial t'} |\Psi_{\alpha,\beta}(t)\rangle. 
\end{equation}}
{where $\Gamma(z)$ is the Euler Gamma function, $\Gamma(z) = \int_0^\infty dt \, t^{z-1} e^{-t}$, with ${\rm Re} \, (z)>0$} \cite{Evangelista2018}. 
This operator is associated with non-Markovian environments, in which a future state depends not only on the present state but also on past states \cite{Wei2024}. 
In this context, $\alpha$ is associated with the degree of memory in the system \cite{Zu2021}.
Since our main interest is to explore the effects of $\alpha$ and $\beta$ in the dynamics {of} the {TD}JC model, we henceforth adopt the notation $a_{\alpha,\beta}(t)$, $b_{\alpha,\beta}(t)$, $W_{\alpha,\beta} (t)$ and $S^A_{\alpha,\beta} (t)$, 
highlighting the dependence on the derivative order and the power of $i$. 

In this scenario, the state \eqref{eq:fock_state} evolves in time according to the TFSE \eqref{eq:ftse} with the Hamiltonian \eqref{eq:int_hamiltonian_jc}, resulting in the following system of coupled differential equations 
\begin{equation}
    \begin{aligned} 
 \label{eqa_frac}
        i^\beta  \frac{\partial^\alpha}{\partial t^\alpha} a_{\alpha,\beta}(t)= {} &\lambda(t)b_{\alpha,\beta}(t)\sqrt{n+1}+\frac{1}{2}\Delta a_{\alpha,\beta}(t), \\ 
        i^\beta  \frac{\partial^\alpha}{\partial t^\alpha} b_{\alpha,\beta}(t)= {} &\lambda(t)a_{\alpha,\beta}(t)\sqrt{n+1}-\frac{1}{2}\Delta b_{\alpha,\beta}(t), 
        \end{aligned}
\end{equation}
assuming $a_{\alpha,\beta}(0) \equiv a_0$ and $b_{\alpha,\beta}(0) \equiv b_0$. 
{It is worth mentioning that the time-dependent coupling $\lambda(t)$ is independent of the fractional parameters $\alpha$ and $\beta$, as it governs distinct physical aspects. 
Section \ref{sec:frac_jc} clarifies the scenarios corresponding to different forms of $\lambda(t)$.}

\section{Time fractional light-matter dynamics}\label{sec:frac_jc}
Having identified the dynamical equations for the time fractional TDJC model, we now explore different time-dependent couplings and analyze how the parameters $\alpha$ and $\beta$ affect the system’s dynamics compared to the standard TDJC model.

\subsection{Constant coupling parameter}
Considering the constant coupling parameter $\lambda(t)=\lambda_0$, the solutions of the system, Eq. \eqref{eqa_frac}, read
\begin{equation}
     \begin{aligned}
    a_{\alpha,\beta}(t)  = {} 
    &
     \frac{a_0}{2} \left[E_\alpha\left(\frac{\Omega_n t^\alpha}{i^\beta}\right)+ E_\alpha\left(\frac{-\Omega_n t^\alpha}{i^\beta}\right)\right] 
 + \left[ \frac{a_0 \left(\frac{\Delta}{2}\right)+b_0\lambda_0 \sqrt{n+1}}{2 \Omega_n} \right] \\
    & \times \left[E_\alpha\left(\frac{\Omega_n t^\alpha}{i^\beta}\right)- E_\alpha\left(\frac{-\Omega_n t^\alpha}{i^\beta}\right)\right],\\
     b_{\alpha,\beta}(t) = {} 
     &
      \frac{b_0}{2} \left[E_\alpha\left(\frac{\Omega_n t^\alpha}{i^\beta}\right)+ E_\alpha\left(\frac{-\Omega_n t^\alpha}{i^\beta}\right)\right] 
 + \left[\frac{a_0\lambda_0 \sqrt{n+1} -b_0 \left(\frac{\Delta}{2}\right)}{2 \Omega_n} \right]\\
    & \times \left[E_\alpha\left(\frac{\Omega_n t^\alpha}{i^\beta}\right)- E_\alpha\left(\frac{-\Omega_n t^\alpha}{i^\beta}\right)\right]. \label{solucao_fracionaria}
    \end{aligned}
\end{equation}
where $E_\alpha(\pm \Omega_n i^{-\beta}t^\alpha)$ is the Mittag-Leffler function \cite{Evangelista2018}, 
which reduces to the exponential function for $\alpha\rightarrow 1$. 
In the same regime, Eq. \eqref{solucao_fracionaria} returns to the solutions found in Eq. \eqref{solucao_inteiro}. 
{
As initial condition, we fix $a_0=1$ and $b_0=0$ with $n=0$, which corresponds to the atom in the excited state and the cavity in the vacuum state.
}
{Atomic trapping \cite{Cirac1990}, a phenomenon arising from superpositions in the atomic degrees of freedom, can mask the time-dependent coupling effects that are the primary focus of this work.
An analysis of the constant coupling case under fractional time evolution, including varying initial photon numbers $n$, is provided in Ref. \cite{Wei2024}.
}

\begin{figure}[t]
\centering
\includegraphics[scale=0.5]{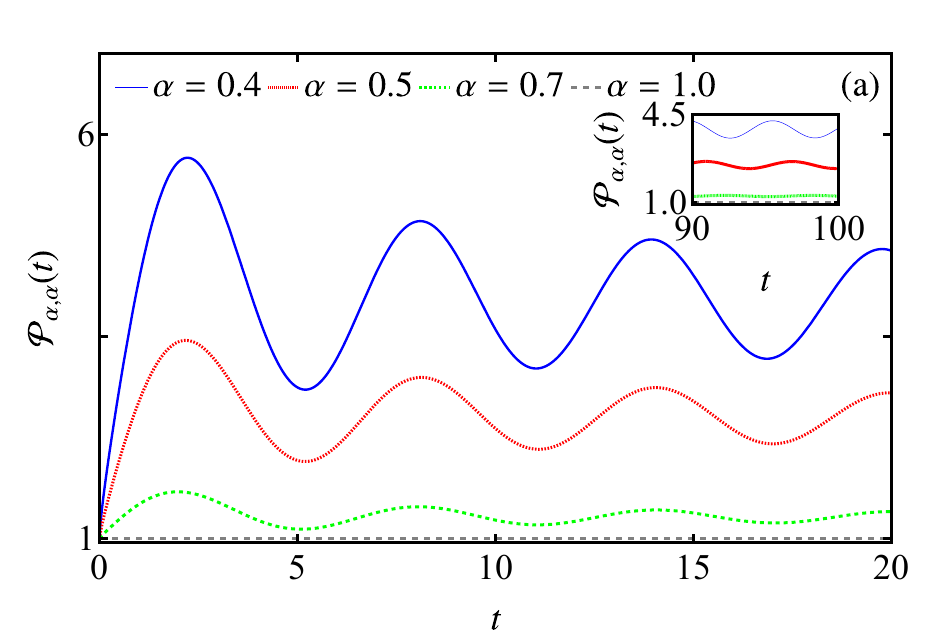}
\hspace{0.5cm}
\includegraphics[scale=0.5]{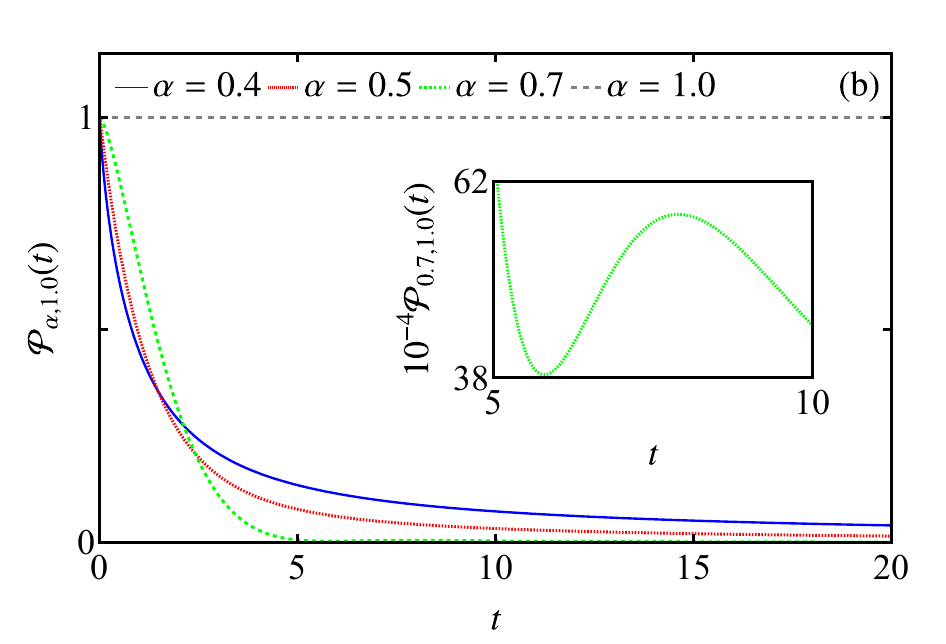}
  \caption{
{Normalization parameter} $\mathcal{P}_{\alpha,\beta}(t)$ as a function of time, setting $a_0=1$, $b_0=0$, $n=0$, $\lambda_0=1$ and $\Delta=0.5$, for different values of $\alpha$: 
$\alpha=0.4$ (solid blue line), $\alpha=0.5$ (finely dotted red line), $\alpha=0.7$ (dotted green line), 
$\alpha=1.0$ (dashed gray line).
In (a), we study the scenario $\alpha=\beta$, while in (b), $\beta=1.0$.
The inset in (a) displays the behavior of $\mathcal{P}_{\alpha,\alpha}(t)$ in the interval $t \in [90,100]$, 
while the inset in (b) exhibits the behavior of $\mathcal{P}_{0.7,1.0}(t)$ in the interval $t \in [5,10]$.
} 
  \label{fig:timev_prob_alpha}
\end{figure}

{
We first analyze the behavior of the normalization parameter $\mathcal{P}_{\alpha,\beta}(t)$, Eq. \eqref{eq:norm}, as it directly influences the system dynamics. 
Its time evolution is shown in Fig \ref{fig:timev_prob_alpha},} assuming $\alpha=0.4$ (solid blue line), $\alpha=0.5$ (finely dotted red line), $\alpha=0.7$ (dotted green line), $\alpha=1.0$ (dashed gray line).
We set $\lambda_0=1$ and $\Delta=0.5$ -- values that will remain fixed for the following plots unless otherwise specified. 
{
In both panels of Fig. \ref{fig:timev_prob_alpha}, it is clear that Eq. \eqref{eq:norm} cannot be interpreted as a probability, since it is not conserved \cite{Laskin2017} (except when $\alpha=\beta=1$) due to the non-unitary evolution associated with the TFSE \cite{Cius2022}.
Consequently, such unusual behavior allows for alternative interpretations of fractional time effects, such as particle creation and annihilation \cite{lu2017,Lu2018}.
Nevertheless, in this work, the quantity $\mathcal{P}_{\alpha,\beta}(t)$ serves solely as a normalization parameter, allowing us to recover the standard quantum mechanical interpretation when analyzing the quantities of interest. 

}
When $\alpha=\beta$ (Fig. \ref{fig:timev_prob_alpha}(a)), we observe a behavior characterized by oscillations around a fixed value with a gradually decreasing amplitude.
Decreasing $\alpha$ results in higher values of $\mathcal{P}_{\alpha,\beta}(t)$, 
while the periodicity in the oscillations remains the same.
For later times, as shown in the inset, the undulatory behavior persists, 
even if the oscillations are very subtle for {$\alpha=0.7$}.
On the other hand, when setting $\beta=1.0$ (Fig. \ref{fig:timev_prob_alpha}(b)), the {normalization parameter} initially decays to a small value near zero and then slowly decreases {over time}.
Furthermore, a decrease in $\alpha$ leads to a faster decay at early times, followed by a slower evolution at later times, resulting in higher values for  $\mathcal{P}_{\alpha,\beta}(t)$. 
This behavior can be interpreted from a dissipative {\cite{Zu2025} perspective}, where information loss occurs due to the flow of information from the system to the environment.
However, as shown in the inset, the behavior is non-monotonic, exhibiting small inflections -- a counterintuitive feature arising from non-Markovian evolution {\cite{Wei2024,Zu2025}}.
{For $\alpha < 0.4$, the normalization parameter evolves similarly in both the $\alpha=\beta$ and $\beta=1$ cases, i.e., the tendency of the dynamical evolution remains, and no new phenomenon is observed.}

The time evolution of $\mathcal{P}_{\alpha,\beta}(t)$ in these two scenarios was first investigated in Ref. \cite{Wei2024}, but considering the on-resonance condition.
In this sense, the influence of detuning and the constant coupling parameter is discussed in the later paragraphs.

\begin{figure}[h]
\centering
\includegraphics[scale=0.5]{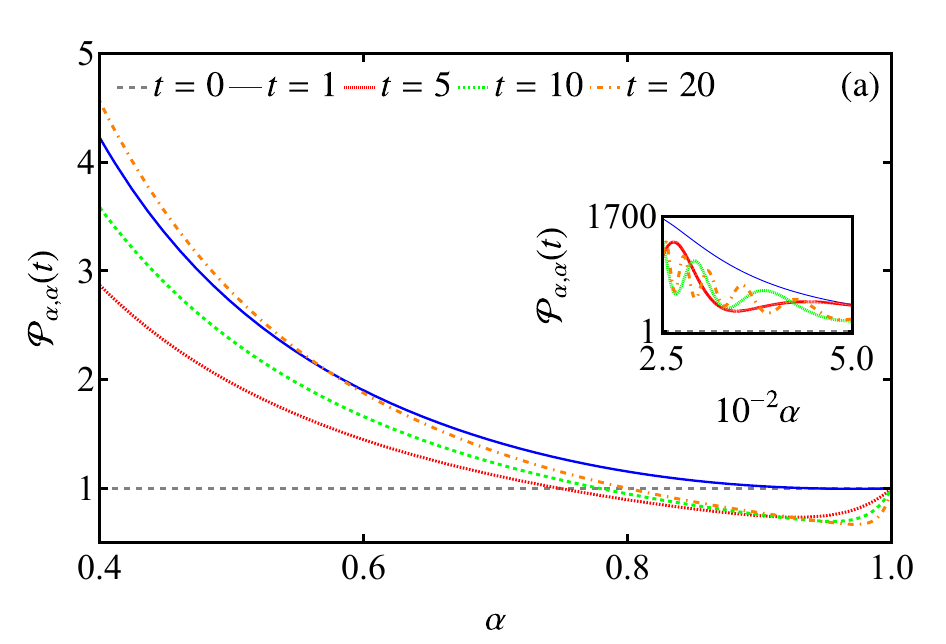}
\hspace{0.5cm}
\includegraphics[scale=0.5]{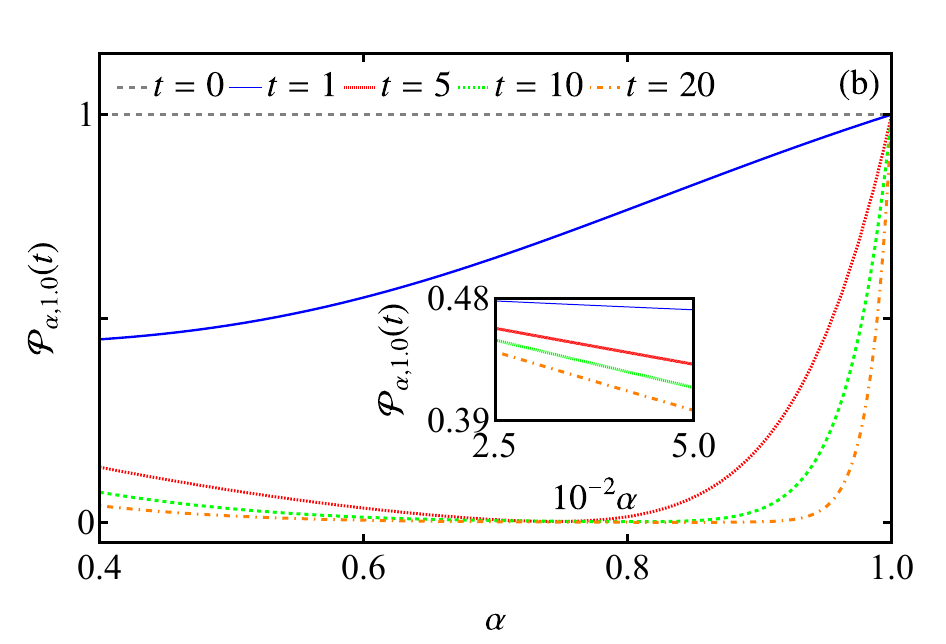}
  \caption{
{Normalization parameter} $\mathcal{P}_{\alpha,\beta}(t)$ as a function of $\alpha$ in the interval $\alpha \in [0.4,1.0]$, setting $a_0=1$, $b_0=0$, $n=0$, $\lambda_0=1$ and $\Delta=0.5$, for different instants: 
$t=0$ (dashed gray line), $t=1$ (solid blue line), $t=5$ (finely dotted red line), $t=10$ (dotted green line), $t=20$ (dot-dashed orange line).
As an inset, we plot the same quantity but in the range $\alpha \in [0.025,0.050]$.
In (a), we study the scenario $\alpha=\beta$, while in (b), $\beta=1.0$.
  } 
  \label{fig:totalprob_alpha}
\end{figure}

For more information on the scope of {
$\mathcal{P}_{\alpha,\beta}(t)$, we study its dependence on $\alpha$ for fixed instants (Fig. \ref{fig:totalprob_alpha}).}
The solid blue line corresponds to $t=1$, the finely dotted red line to $t=5$, the dotted green line to $t=10$, the dot-dashed orange line to $t=20$, and the dashed gray line to $t=0$. 
Naturally, when $t=0$, we have $\mathcal{P}_{\alpha,\beta}(t)=1$ for all values of $\alpha$ in both scenarios ($\alpha=\beta$ and $\beta=1.0$).
When $\alpha=\beta$ (Fig. \ref{fig:totalprob_alpha}(a)), the {normalization parameter} may reach values below one at later times ($t>1$), particularly for higher values of $\alpha$.
For $\alpha\ll1.0$, exceptionally large values of {$\mathcal{P}_{\alpha,\beta}(t)$} can be obtained, 
as displayed in the inset. 
In contrast, for $\beta=1.0$, losses are observed (Fig. \ref{fig:totalprob_alpha}(b) and the inset therein), 
although the values are restricted to $\mathcal{P}_{\alpha,\beta}(t) \leq1.0$, for arbitrary $\alpha$.

\begin{figure}[h]
\centering
\includegraphics[scale=0.5]{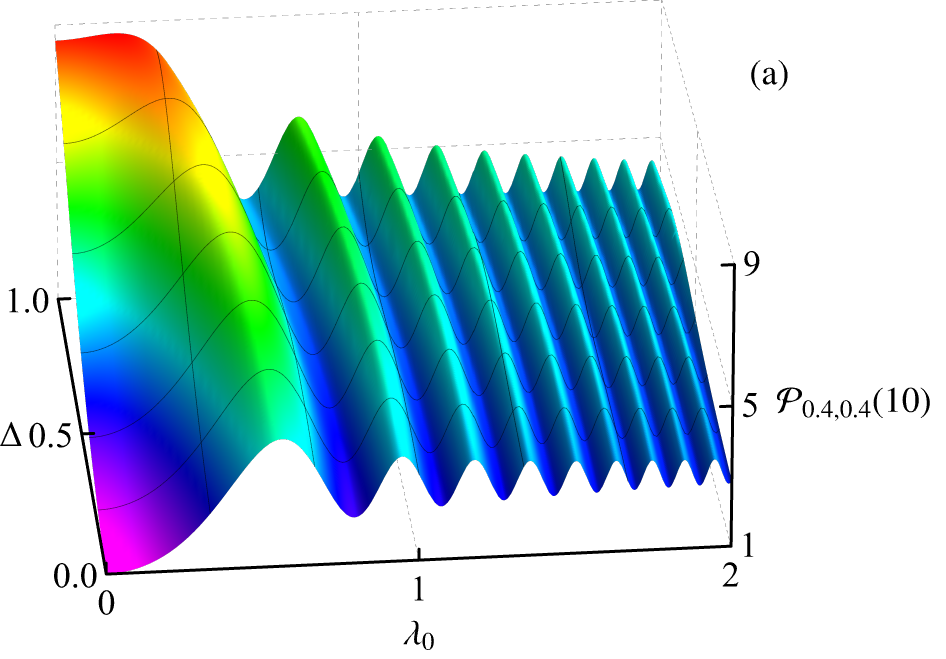}
\hspace{0.5cm}
\includegraphics[scale=0.5]{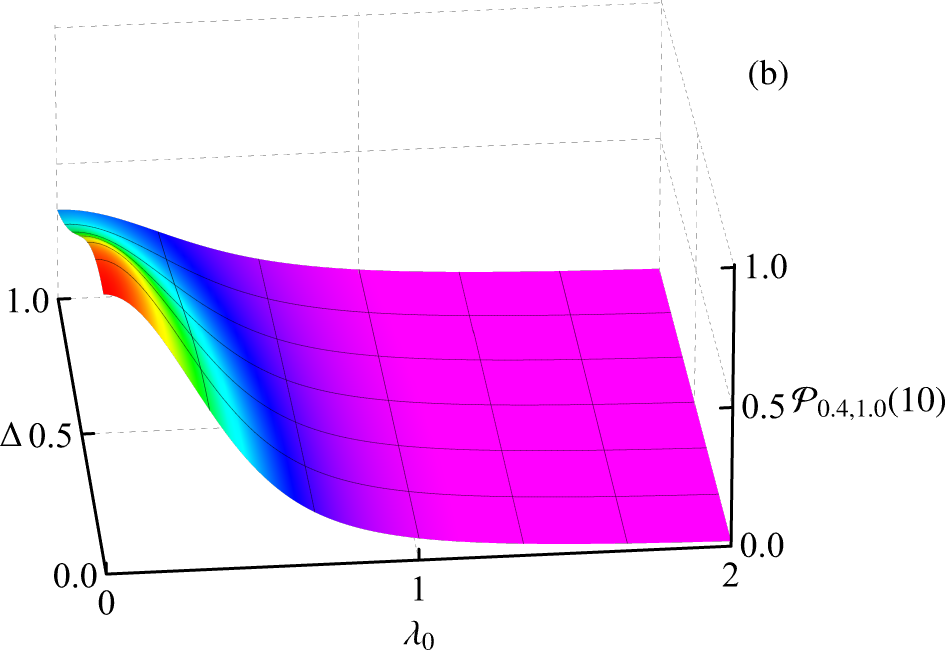}
  \caption{
The quantity $\mathcal{P}_{\alpha,\beta}(t)$ as a function of the parameters $\lambda_0$ and $\Delta$, for $t=10$, $\alpha =0.4$, $a_0=1$, $b_0=0$ and $n=0$. 
Panel (a) is for $\beta=\alpha$, and panel (b) is for $\beta=1.0$. } 
  \label{fig:plot3d_totalprob}
\end{figure}

The choice of $\lambda_0$ and $\Delta$ { affects} the behavior of the {normalization parameter} (Fig. \ref{fig:plot3d_totalprob}).
In the case of $\alpha=\beta$ (Fig. \ref{fig:plot3d_totalprob}(a)), larger detuning favors the emergence of higher values of $\mathcal{P}_{\alpha,\beta}(t)$, significantly when $\lambda_0<1$ \cite{Wei2024}{.}
The dependence on $\lambda_0$ exhibits an oscillatory behavior; this results from the fact that $\lambda_0$ modifies both the rate at which the oscillations of $\mathcal{P}_{\alpha,\beta}(t)$ occur and their amplitude.
Generally, a smaller $ \lambda_0 $ {  implies} in higher values of the {normalization parameter}.
Particularly, the case when $\lambda_0=0$ (effectively uncoupling the atom and the cavity mode) was studied in Ref. \cite{Lu2018}. 
When the cavity is initially in a Fock state {$(n\neq0)$}, the comments made regarding $\lambda_0$ can be extended to the term $\sqrt{n+1}$.

The opposite happens for $\beta=1.0$ (Fig. \ref{fig:plot3d_totalprob}(b)), where $\mathcal{P}_{\alpha,1.0}(t) \rightarrow 1$ as $(\Delta,\lambda_0) \rightarrow0$. 
In addition, we observe that for $\lambda_0 > 1$, the dynamics are independent of $\Delta$, and dissipative effects {prevail}, causing a rapid decay of the {normalization parameter}.
Depending on the choice of $\lambda_0$, the minimum value of $\mathcal{P}_{\alpha,1.0}(t)$ can be reached earlier. 

{As a consequence of of the non-conservation of probability, the direct calculation of observables from $|\Psi_{\alpha,\beta}(t)\rangle$
suggests some physical divergences from the standard formulation \cite{lu2017,Lu2018}.}
In this sense, to statistically characterize the behavior of the quantities of interest, we employ a normalization of the density matrix and the population inversion \cite{SERGI2013,Zu2021}, which follows 
\begin{equation} \label{eq:normalization}
    \begin{aligned}
        \hat{\rho}^\prime_{\alpha,\beta}(t) & =\frac{\hat{\rho}_{\alpha,\beta}(t)}{\mathcal{P}_{\alpha,\beta}(t)}, \\
        W_{\alpha,\beta}^\prime(t)&=\frac{W_{\alpha,\beta}(t)}{\mathcal{P}_{\alpha,\beta}(t)}.
    \end{aligned}
\end{equation}
{Equivalently, this procedure corresponds to renormalizing the state
$|\Psi_{\alpha,\beta}(t)\rangle \mapsto |\Psi^\prime_{\alpha,\beta}(t)\rangle$
and computing expectation values with a trace-one density operator, as customary for non-unitary dynamics \cite{Dalibard1992,BreuerPetruccione2002,SERGI2013}.} 
The normalized VNE ($S_{\alpha,\beta}^{A\prime}(t)$)  follows from $\hat{\rho}^\prime_{\alpha,\beta}(t)$. 
Assuming the initial state {$|\Psi_{\alpha,\beta}(0)\rangle =|e,n\rangle $}, the equations for the normalized population inversion and VNE are
\begin{equation}
  \begin{aligned}
W^\prime_{\alpha,\beta}(t)= {} 
          & \frac{1}{\mathcal P_{\alpha,\beta}(t)}\left[|a_{\alpha,\beta}(t)|^2-|b_{\alpha,\beta}(t)|^2\right], \\
         S^{A\prime}_{\alpha,\beta}(t) = {} 
         & -\frac{1}{\mathcal{P}_{\alpha,\beta}(t)} 
\left[ |a_{\alpha,\beta}(t)|^2 \log_2 \left( \frac{|a_{\alpha,\beta}(t)|^2}{\mathcal{P}_{\alpha,\beta}(t)} \right) \right.
\\ 
&+ \left. |b_{\alpha,\beta}(t)|^2 \log_2 \left( \frac{|b_{\alpha,\beta}(t)|^2}{\mathcal{P}_{\alpha,\beta}(t)} \right) \right].
    \end{aligned}  
\end{equation}
We perform this normalization to resemble the interpretation of the standard formulation, but this procedure does not eliminate the non-Markovian effects due to the FC formulation \cite{Naber2004,Iomin2009}.
It is worth noting that, within the considered time window and given the choice of $\alpha$, the amount $\mathcal{P}_{\alpha,\beta}(t)$ does not {cause divergences}.
From this point onward, we will omit the term ``normalized'' when referring to the normalized population inversion and  VNE.

\begin{figure}[t]
\centering
\includegraphics[scale=0.5]{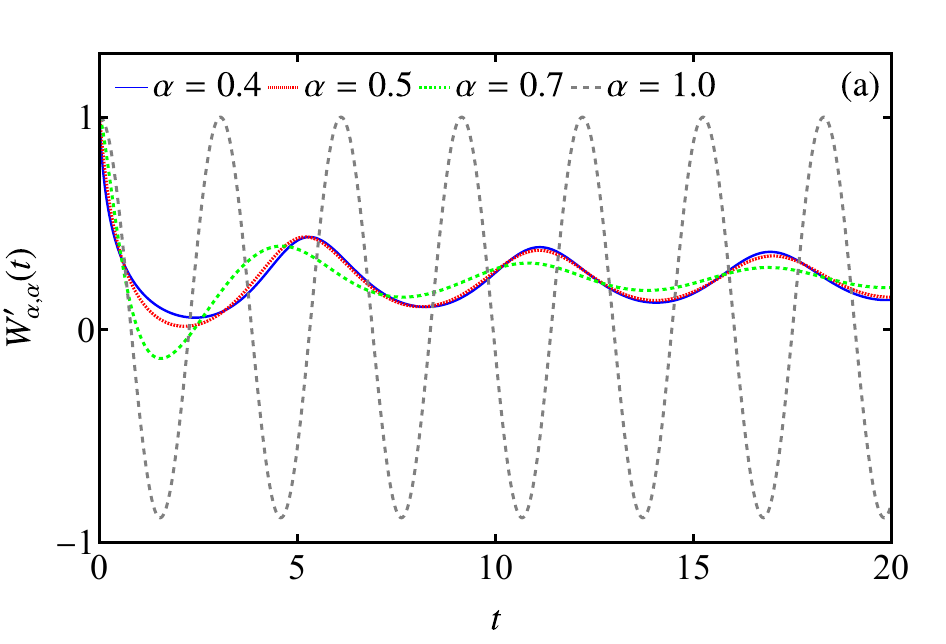}
\hspace{0.5cm}
\includegraphics[scale=0.5]{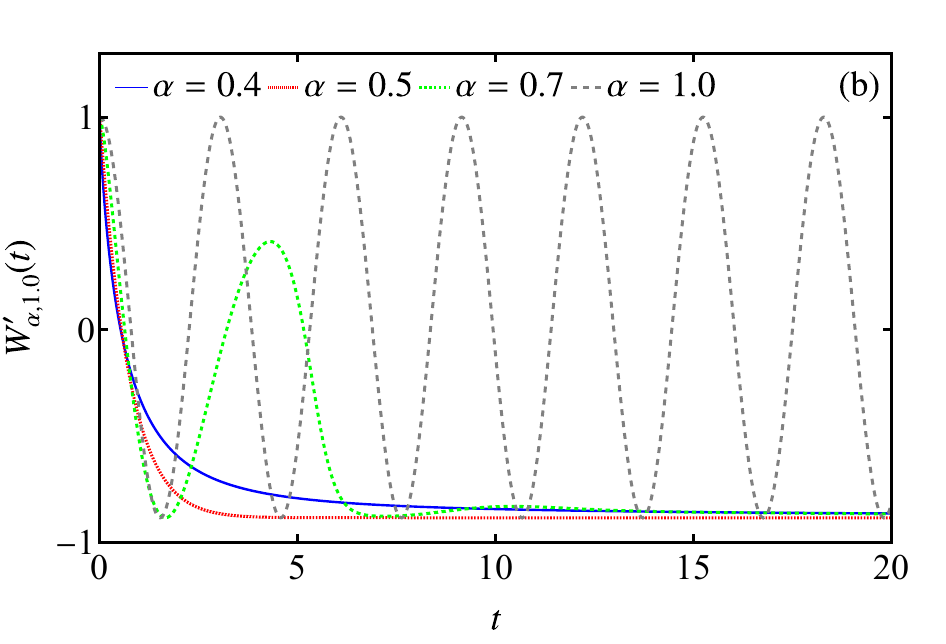}
  \caption{
The normalized population inversion resulting from a constant coupling parameter as a function of $t$, setting $a_0=1$, $b_0=0$, $n=0$, $\lambda_0=1$ and $\Delta=0.5$, for different $\alpha$: 
$\alpha=0.4$ (solid blue line), $\alpha=0.5$ (finely dotted red line) and $\alpha=0.7$ (dotted green line), $\alpha=1.0$ (dashed gray line). 
Panel (a) is for $\beta=\alpha$, and panel (b) is for $\beta=1.0$.
  } 
  \label{fig:popinv1}
\end{figure}

The population inversion when $\lambda(t)=\lambda_0$ is shown in Fig. \ref{fig:popinv1}, where the panel (a) is for $\alpha=\beta$ and (b) is for $\beta=1.0$. 
For $\alpha=1.0$, the gray dashed line, we recover the standard vacuum RO, 
whose behavior is well known \cite{GERRY2005}. 
However, for $\alpha \in (0,1)$, 
we observe a diminishing in the amplitudes as well as a change in the corresponding periods, 
causing  $W^\prime_{\alpha,\beta}(t)$ to oscillate around a fixed value for longer times. 
Furthermore, asymmetry around the $x$ axis can be attributed to the off-resonance condition.
The oscillations are suppressed by setting $\beta=1.0$, as observed in Fig. \ref{fig:popinv1}(b). For $\alpha=0.7$ (dotted green line), there are initial oscillations -- a consequence of the small inflections in {$\mathcal{P}_{\alpha,\beta}(t)$} -- before the population inversion goes to an asymptotic value. 
This behavior, in which the JC interaction initially develops oscillations before the FC intervention, can be interpreted as an intermediate case between the trivial scenario and the one corresponding to smaller values of $\alpha$.
By further decreasing $\alpha$, the solution is just a decay in $W^\prime_{\alpha,1.0}(t)$, influenced by a dissipative environment.

For the initial state considered, the VNE depends only on the square modulus of $a_{\alpha,\beta}(t)$ and $b_{\alpha,\beta}(t)$. 
Therefore, atom-field entanglement can be understood in terms of atomic populations.
For example, when $|W^\prime_{\alpha,\beta}(t)|=1$, we have $S^{A\prime}_{\alpha,\alpha}(t)=0$, because the atom is, surely, at the excited or ground state, and the state, Eq. \eqref{eq:fock_state}, is separable. 
On the other hand, when $W^\prime_{\alpha,\beta}(t)=0$, the entropy reaches its maximum value since the atomic subsystem is maximally mixed -- reflecting equal probabilities in the joint atom-field state.

\begin{figure}[h]
\centering
\includegraphics[scale=0.5]{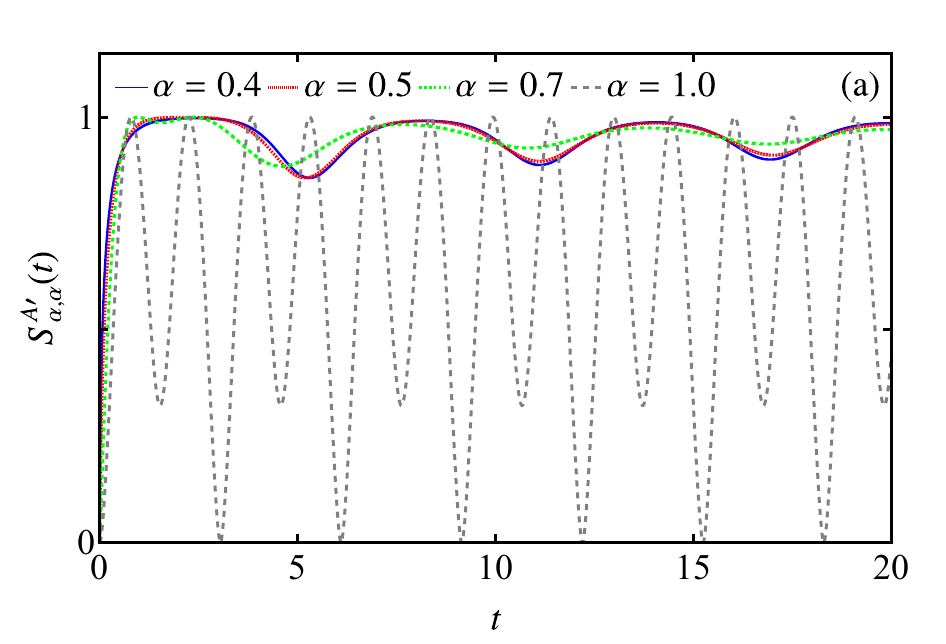}
\hspace{0.5cm}
\includegraphics[scale=0.5]{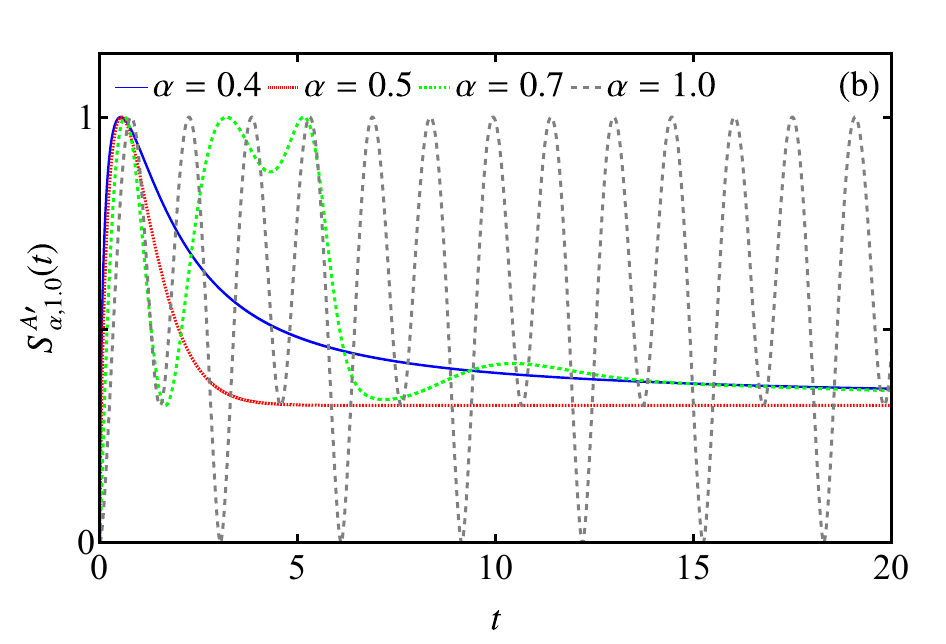}
  \caption{
The normalized VNE for a constant coupling parameter as a function of $t$, setting $a_0=1$, $b_0=0$, $n=0$, $\lambda_0=1$ and $\Delta=0.5$, for different $\alpha$: 
$\alpha=0.4$ (solid blue line), $\alpha=0.5$ (finely dotted red line) and $\alpha=0.7$ (dotted green line), $\alpha=1.0$ (dashed gray line). 
Panel (a) is for $\beta=\alpha$, and panel (b) is for $\beta=1.0$.
  } 
  \label{fig:vne1}
\end{figure}

For the constant coupling, the VNE is presented in Fig. \ref{fig:vne1}. 
In the case of $\alpha=\beta$ (Fig. \ref{fig:vne1}(a)), we observe a relatively high entanglement, since
the population inversion fluctuates around a value closer to zero. 
In this sense, FC contributes to the mixedness of the atomic subsystem. 
For $\beta=1.0$ (Fig. \ref{fig:vne1}(b)),
we obtain asymptotic values due to the decay in the interaction.
The asymptotic values are lower than in the $\alpha =\beta$ case because the interaction ceases when $W^\prime_{\alpha,\beta}(t)$ is close to $-1$.
The effects of detuning become evident when compared to the on-resonance framework of Ref. \cite{Zu2025}, where, even for $\alpha = 0.4$, residual oscillations were still observed before the onset of an exponential-like asymptotic behavior.

\subsection{Linear coupling}
The linear coupling modulation, given by
\begin{equation} \label{eq:coup_lin}
    \lambda(t) = \lambda_0(\zeta t),
\end{equation}
was first considered by Joshi and Lawande \cite{Joshi1993}.
This time dependence can be used to model the scenario of 
a well-localized atom in a varying mode \cite{LARSON2021}. 
In this framework, the control parameters determine whether the {mode} change is adiabatic or sudden. 

From now on, our goal is to study the modifications into $W^\prime_{\alpha,\beta}(t)$ and  $S^{A\prime}_{\alpha,\beta}(t)$ induced by 
different formats of $\lambda(t)$. 
In this way, a closed solution as previously derived becomes challenging for arbitrary $\lambda(t)$. 
To solve this problem, we resort to numerical integration of Eqs. \eqref{eqa_frac} for a prescribed $\lambda(t)$. 
In this subsection, $\lambda(t)$  is defined by Eq. \eqref{eq:coup_lin}. 
Throughout this work, the numerical method employed is the Adams-Bashforth-Moulton method, as presented by Diethelm, Ford, and Freed \cite{diethelm2005}, with a step size of $0.001$. 

\begin{figure}[t]
\centering
\includegraphics[scale=0.5]{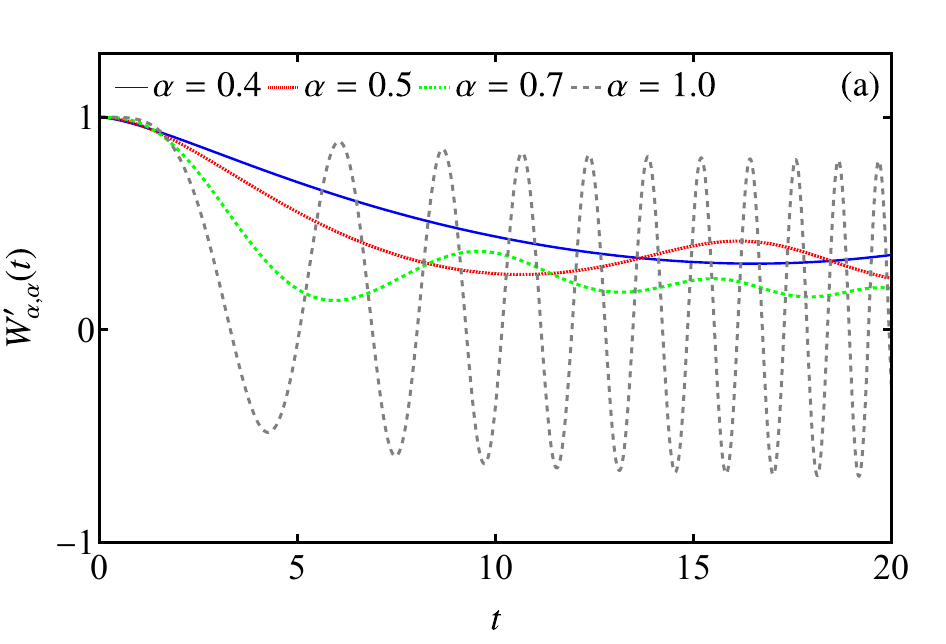}
\hspace{0.5cm}
\includegraphics[scale=0.5]{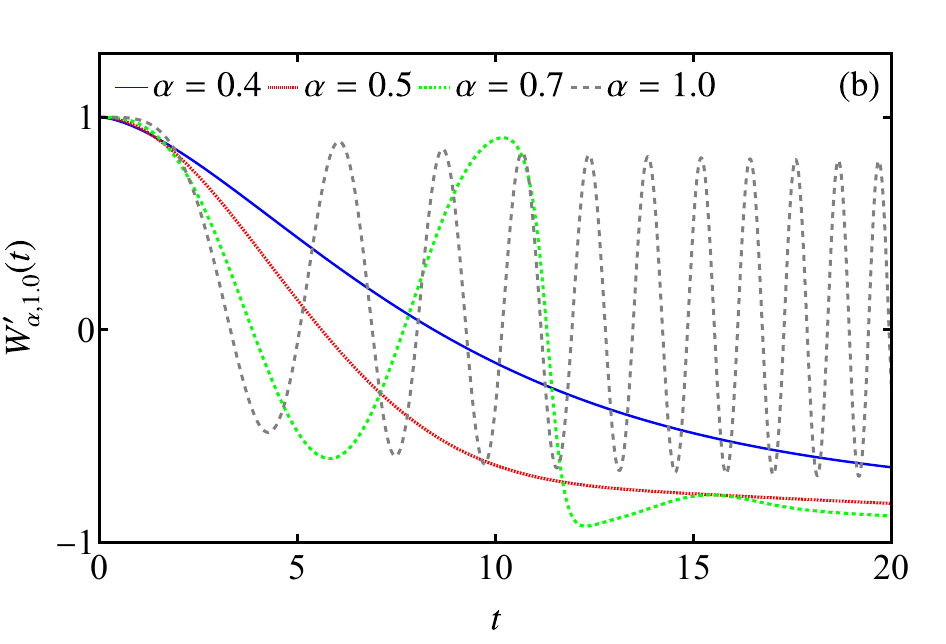}
  \caption{
The normalized population inversion resulting from a linear coupling parameter (Eq. \eqref{eq:coup_lin}) as a function of $t$, setting $a_0=1$, $b_0=0$, $n=0$, $\lambda_0=1$, $\Delta=0.5$, and $\zeta=0.16$ for different $\alpha$: 
$\alpha=0.4$ (solid blue line), $\alpha=0.5$ (finely dotted red line) and $\alpha=0.7$ (dotted green line), $\alpha=1.0$ (dashed gray line). 
Panel (a) is for $\beta=\alpha$, and panel (b) is for $\beta=1.0$. 
  } 
  \label{fig:lin_popinv}
\end{figure}

Setting $\zeta = 0.16$, we model a sudden change scenario.
In this case, Fig. \ref{fig:lin_popinv} displays the evolution of $W^\prime_{\alpha,\alpha}(t)$ in panel (a) and $W^\prime_{\alpha,1.0}(t)$ in  (b). 
In the standard case $\alpha=\beta=1.0$ (the gray dashed line in both plots), RO suffer acceleration, with amplitudes that initially increase.
The oscillations near $t = 0$ are slower because $\lambda(0) = 0$, 
resulting in a weaker interaction and a less efficient exchange between the atom and the cavity.
For values of $\alpha < 1$, we note effects similar to those observed in the constant coupling scenario, but now with an extended period in the oscillations and a further upward displacement along the $y$-axis.
The case of $\beta=1.0$ (Fig. \ref{fig:lin_popinv}(b)) results in a contrasting behavior.
For example, for $\alpha = 0.7$ (dotted green line) and $\alpha = 0.5$ (finely dotted red line), we obtain values closer to $-1$ at {later times} than in the case with $\alpha = \beta = 1.0$.
Compared to the case with $\lambda(t) = \lambda_0$, $W^{\prime}_{\alpha,\beta}(t)$ also reaches a stationary value, but the time at which this occurs now depends on $\zeta$ as well as $\lambda_0$.

\begin{figure}[t]
\centering
\includegraphics[scale=0.5]{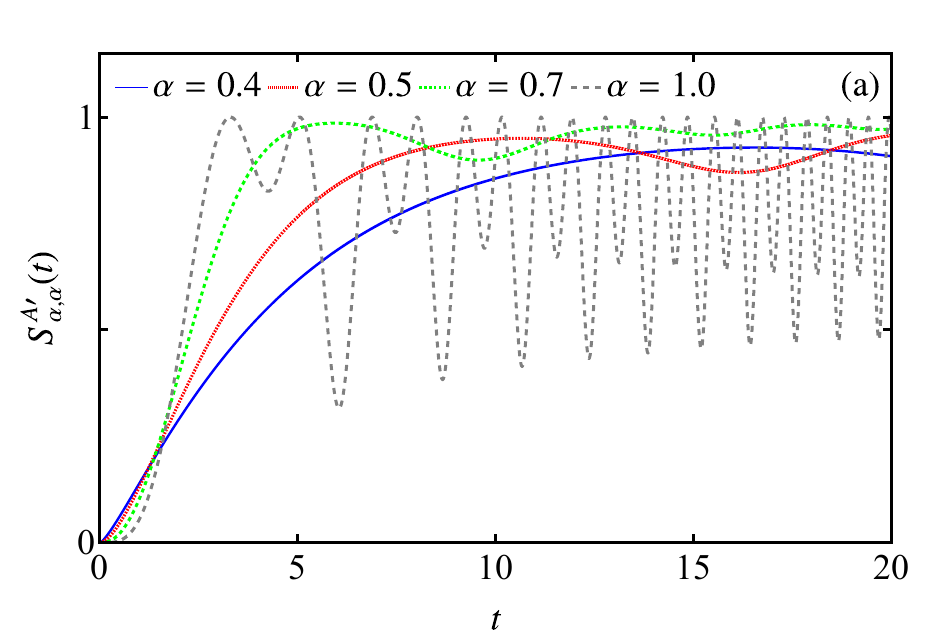}
\hspace{0.5cm}
\includegraphics[scale=0.5]{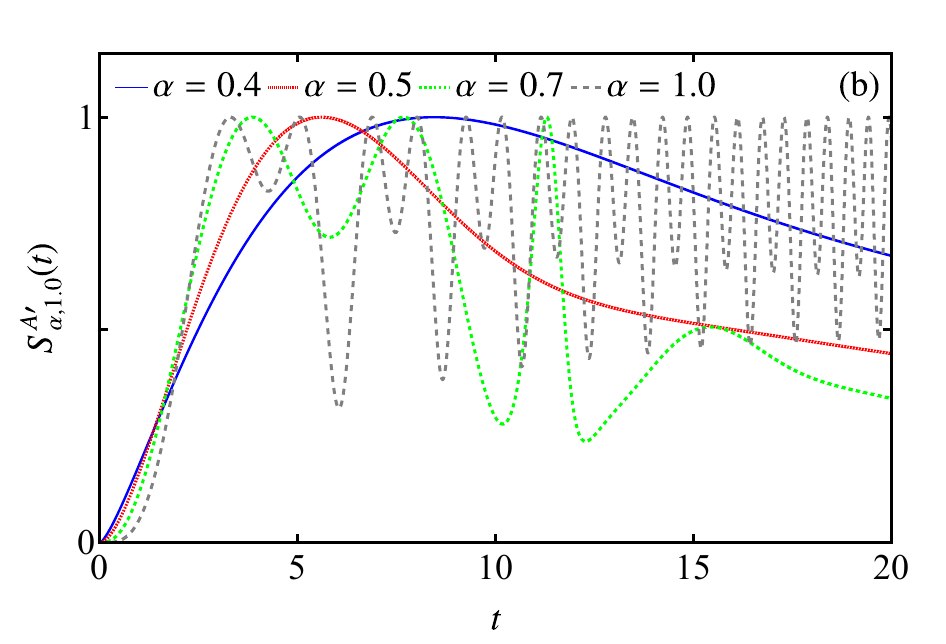}
  \caption{
The normalized VNE resulting from a linear coupling parameter (Eq. \eqref{eq:coup_lin}) as a function of $t$, setting $a_0=1$, $b_0=0$, $n=0$, $\lambda_0=1$, $\Delta=0.5$, and $\zeta=0.16$ for different $\alpha$: 
$\alpha=0.4$ (solid blue line), $\alpha=0.5$ (finely dotted red line) and $\alpha=0.7$ (dotted green line), $\alpha=1.0$ (dashed gray line). 
Panel (a) is for $\beta=\alpha$, and panel (b) is for $\beta=1.0$. 
  } 
  \label{fig:linvne1}
\end{figure}

The VNE (Fig. \ref{fig:linvne1}) in the standard case (dashed gray line) follows intuitively from the behavior of the population inversion: the oscillations it undergoes have the period amplified by the accelerated RO.
In both scenarios, $\alpha=\beta$ (Fig. \ref{fig:linvne1}(a)) and $\beta=1.0$ (Fig. \ref{fig:linvne1}(b)), the behavior can be readily understood from the corresponding lines for population inversion, and the interpretation is derived from the constant coupling scenario.

\subsection{Exponential coupling}
Prants and Yacoupova \cite{Prants1992} explored analytical solutions from the perspective of the Nikitin model \cite{Nikitin1984}, 
considering a coupling given by 
\begin{equation}\label{exp_coupling}
    \lambda(t) = \lambda_0 e^{\zeta t}.
\end{equation}
This formulation can be phenomenologically interpreted as a model of transient effects in the cavity. 

\begin{figure}[t]
\centering
\includegraphics[scale=0.5]{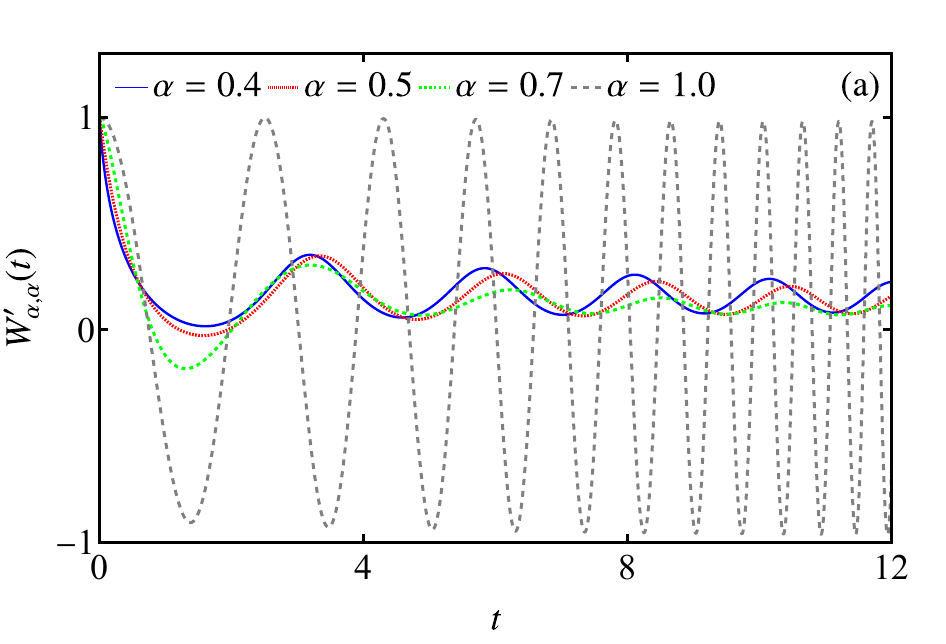}
\hspace{0.5cm}
\includegraphics[scale=0.5]{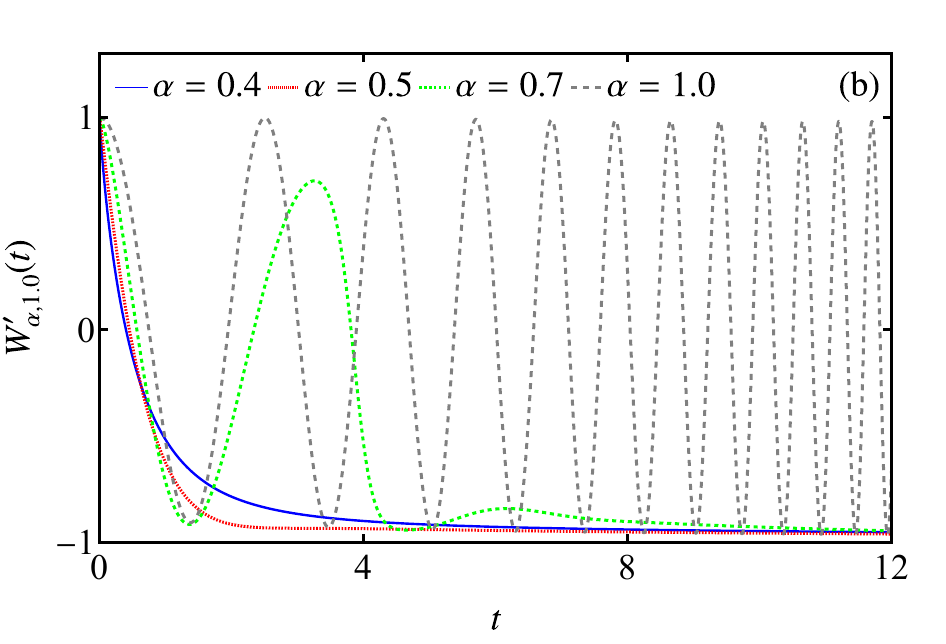}
  \caption{
The normalized population inversion resulting from an exponential coupling parameter (Eq. \eqref{exp_coupling}) as a function of $t$, setting $a_0=1$, $b_0=0$, $n=0$, $\lambda_0=1$, {$\zeta=0.16$} and $\Delta=0.5$, for different $\alpha$: 
$\alpha=0.4$ (solid blue line), $\alpha=0.5$ (finely dotted red line) and $\alpha=0.7$ (dotted green line), $\alpha=1.0$ (dashed gray line). 
Panel (a) is for $\beta=\alpha$, and panel (b) is for $\beta=1.0$.
  } 
  \label{fig:exp_popinv}
\end{figure}

\begin{figure}[t]
\centering
\includegraphics[scale=0.5]{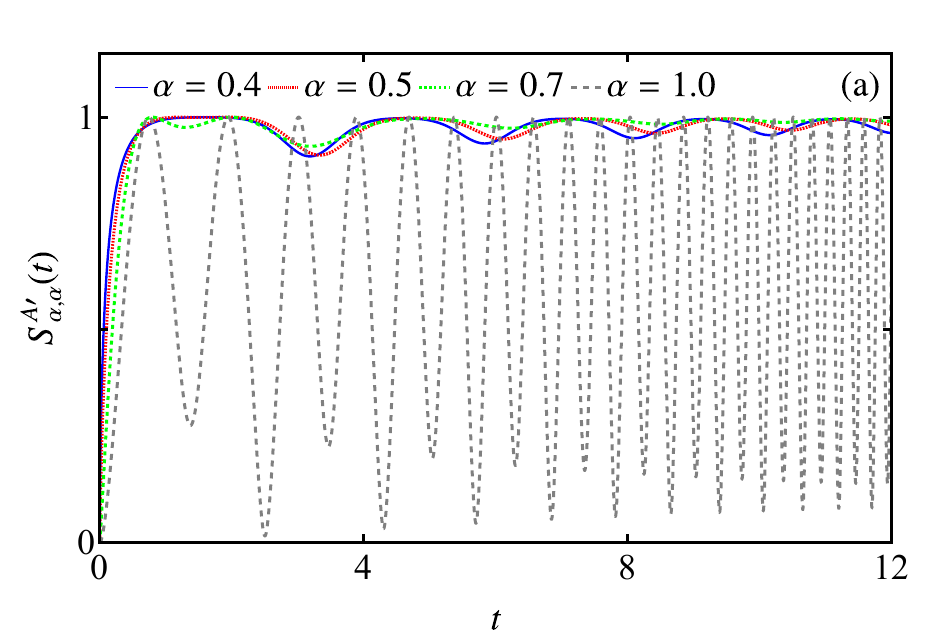}
\hspace{0.5cm}
\includegraphics[scale=0.5]{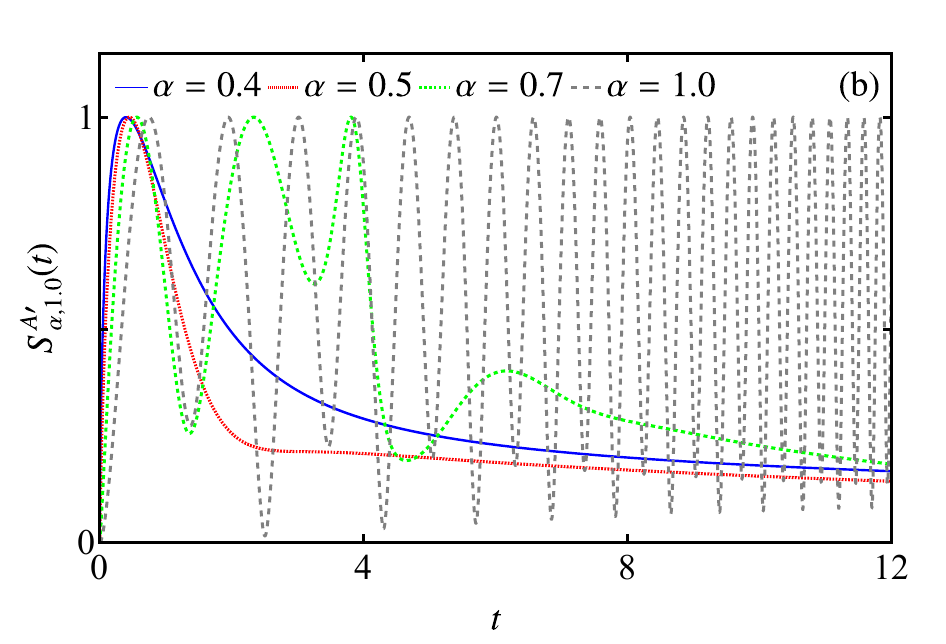}
  \caption{
The normalized VNE resulting from an exponential coupling parameter (Eq. \eqref{exp_coupling}) as a function of $t$, setting $a_0=1$, $b_0=0$, $n=0$, $\lambda_0=1$, {$\zeta=0.16$}  and $\Delta=0.5$, for different $\alpha$: 
$\alpha=0.4$ (solid blue line), $\alpha=0.5$ (finely dotted red line) and $\alpha=0.7$ (dotted green line), $\alpha=1.0$ (dashed gray line). 
Panel (a) is for $\beta=\alpha$, and panel (b) is for $\beta=1.0$.
  } 
  \label{fig:expvne1}
\end{figure}

Figure \ref{fig:exp_popinv} exhibits the time evolution of $W^\prime_{\alpha,\alpha}(t)$ in (a) and $W^\prime_{\alpha,1.0}(t)$ in (b), 
employing $\zeta=0.16$.
In both cases, the shape of the solutions differs from that obtained with the linear coupling but resembles the shape associated with the constant coupling. 
Comparing the results from the constant coupling (Fig. \ref{fig:popinv1}) and $\alpha=\beta$ with those for the exponential modulation, we observe a clear difference in the oscillation period and a subtle variation in the amplitudes over time.  
In this sense, the exponential coupling leads to a progressively shorter period and a more pronounced difference than in the linear case, because the exponential term is nonzero from the outset, resulting in a stronger interaction and enhanced exchange of quanta among the system's degrees of freedom.
When $\alpha<1.0$, this coupling parameter also gives rise to faster dynamics.
The effects are similar when $\beta = 1.0$ (Fig. \ref{fig:exp_popinv}(b)), which is quite similar to Fig. \ref{fig:popinv1}(b), 
but with a faster decay.
The comments regarding population inversion can be extended to the VNE (Fig. \ref{fig:expvne1}), with the additional detail: the asymptotic values for $\beta = 1.0$ differ from those in the constant {and linear} coupling scenario{s}, as the RO cease at different times.

\subsection{Sinusoidal coupling}
In cavity QED, the atom traverses the cavity. 
However, in the case of constant coupling, the atomic motion is disregarded.
Incorporating it, under the assumption that the kinetic energy is significantly greater than the interaction energy \cite{LARSON2021}, leads to a time-dependent coupling modeled as
\begin{equation}\label{sin_coupling}
    \lambda(t)=\lambda_0 \sin(p \zeta t),
\end{equation}
where $p$ is the number of half-wavelengths of the field mode. 
In this scenario, $\zeta$ is a combination of the atomic velocity $v$ and
cavity length $L$, such that $\zeta=v \pi/L$.  
This approach was first proposed by Schlicher \cite{Schlicher1989}, 
and results in a periodic population inversion \cite{Fang1998}. 
Particularly interesting is the fact that when { trigonometric modulations}, such as Eq. \eqref{sin_coupling}, are considered{,} dynamical chaos can emerge for a given set of parameters \cite{Pomeau1986,Prants1997}.
{This occurs because including Eq. \eqref{sin_coupling} in the differential equation system (Eq. \eqref{eqa_frac}) renders it non-autonomous, thereby increasing the dimension of the system's phase space.
A more detailed analysis of the resulting dynamics, including Lyapunov exponents for various parameters, is provided in Ref. \cite{Prants1997}.}
In this section, we set $\zeta=p=1.0$.

\begin{figure}[t]
\centering
\includegraphics[scale=0.5]{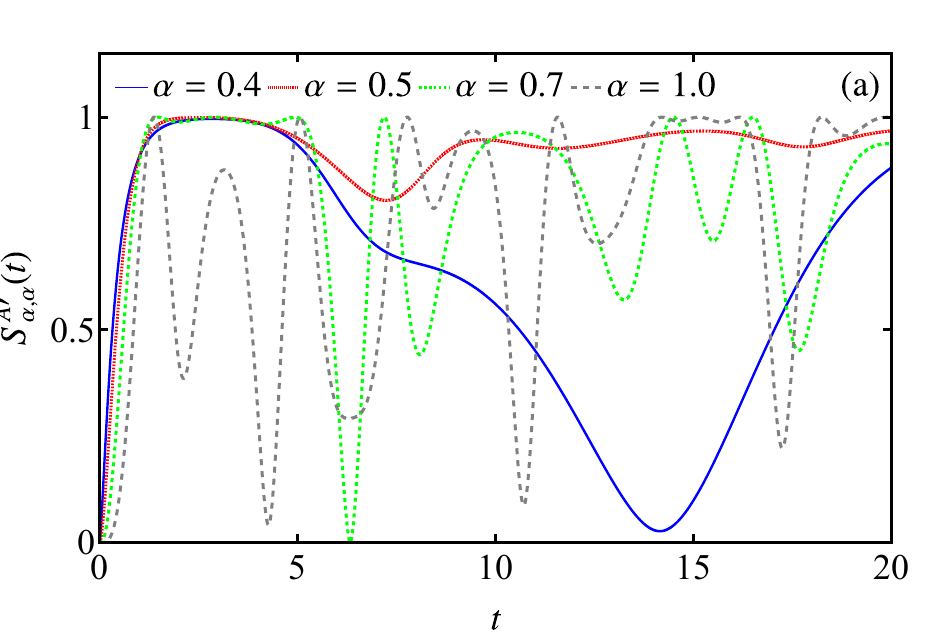}
\hspace{0.5cm}
\includegraphics[scale=0.5]{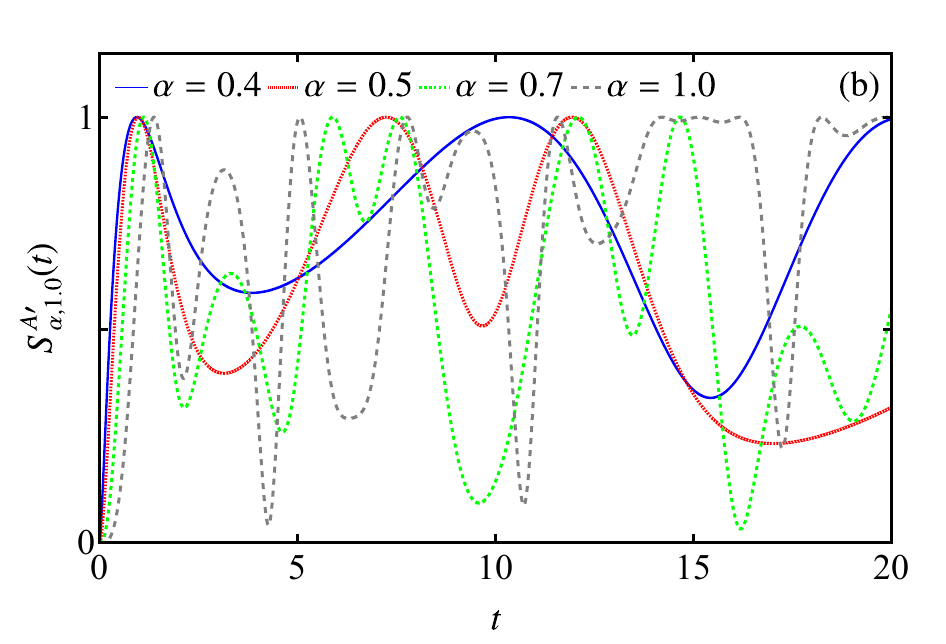}
  \caption{
The normalized VNE resulting from an {sinusoidal} coupling parameter (Eq. \eqref{sin_coupling}) as a function of $t$, setting $a_0=1$, $b_0=0$, $n=0$, $\lambda_0=1$, $\zeta=p=1.0$ and $\Delta=0.5$, for different $\alpha$: 
$\alpha=0.4$ (solid blue line), $\alpha=0.5$ (finely dotted red line) and $\alpha=0.7$ (dotted green line), $\alpha=1.0$ (dashed gray line). 
Panel (a) is for $\beta=\alpha$, and panel (b) is for $\beta=1.0$.
  } 
  \label{fig:sinvne1}
\end{figure}

Compared with the previous results, a coupling given by Eq. \eqref{sin_coupling} leads to very different dynamics for both cases, $W^\prime_{\alpha,\alpha}(t)$ (Fig. \ref{fig:sin_popinv}(a)) and {$W^\prime_{\alpha,1.0}(t)$} (Fig. \ref{fig:sin_popinv}(b)).
The solutions for $W^\prime_{\alpha,\alpha}(t)$ are non-periodic, for all values of $\alpha$.
It is noteworthy that, when $\alpha = 0.4$ {(solid blue line)}, the curve reaches values close to {$W'_{\alpha,\alpha}(t) = 1$}.
In particular, for $\alpha = 0.7$ {(dotted green line)}, the population inversion reaches $-1$ when $t$ is near $6$.
Contrary to the results we have obtained so far, setting $\beta = 1.0$ does not drive the system to a steady state. 
On the other hand, the solutions remain oscillating, being non-periodic for $\alpha=0.7$ and presenting a certain regularity for $\alpha=0.4$ and $\alpha=0.5$ {(finely dotted red line).

\begin{figure}[t]
\centering
\includegraphics[scale=0.5]{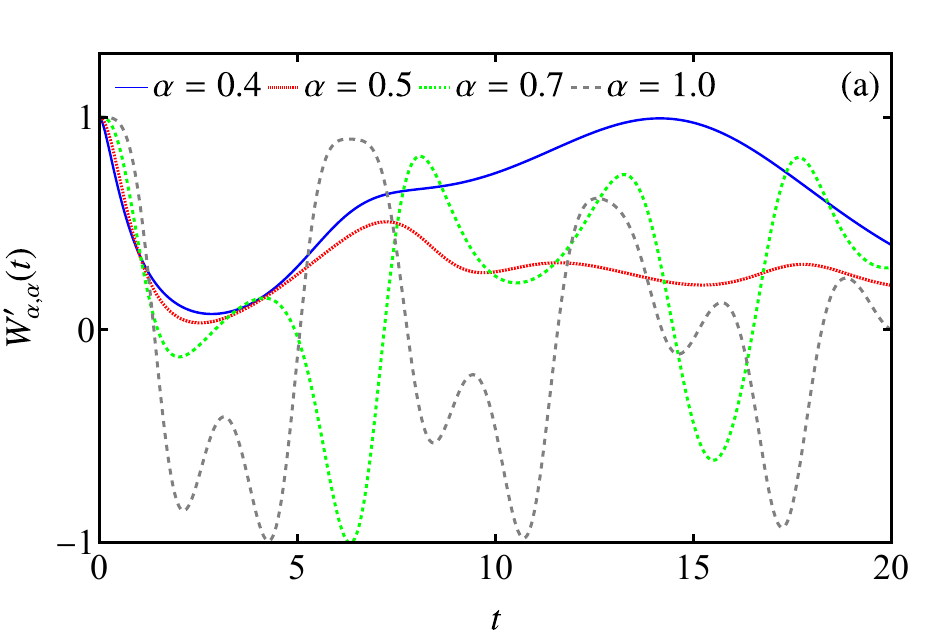}
\hspace{0.5cm}
\includegraphics[scale=0.5]{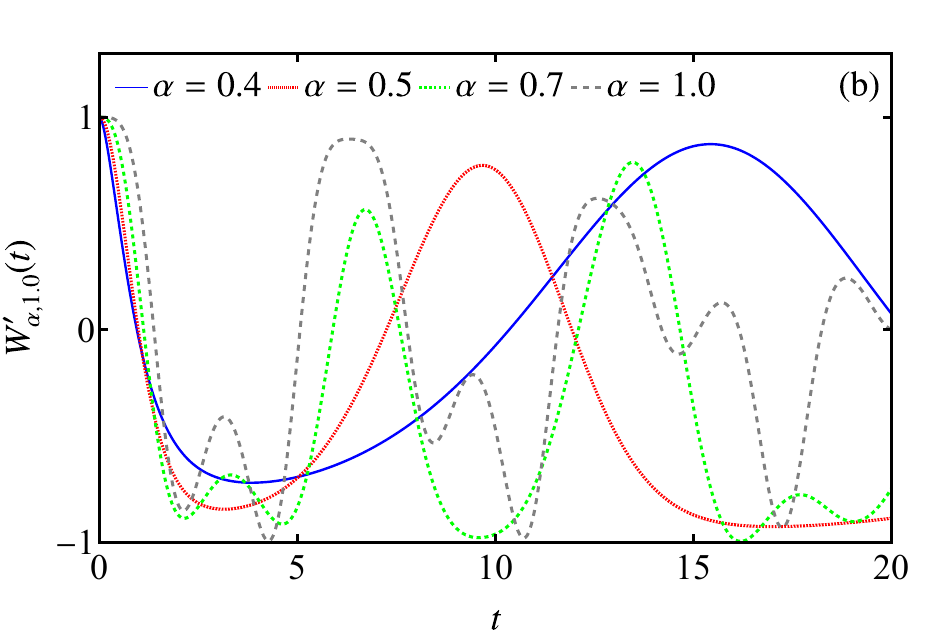}
  \caption{
Population inversion for sinusoidal coupling parameter (Eq. \eqref{sin_coupling}) as a function of $t$, setting $a_0=1$, $b_0=0$, $n=0$, $\lambda_0=1$, $\zeta=p=1.0$ and $\Delta=0.5$, for different $\alpha$: 
$\alpha=0.4$ (solid blue line), $\alpha=0.5$ (finely dotted red line) and $\alpha=0.7$ (dotted green line), $\alpha=1.0$ (dashed gray line). 
Panel (a) is for $\beta=\alpha$, and panel (b) is for $\beta=1.0$.
  } 
  \label{fig:sin_popinv}
\end{figure}

The non-periodic motion is also observed in $S^{A\prime}_{\alpha,\alpha}(t)$ (Fig. \ref{fig:sinvne1}(a)) and  $S^{A\prime}_{\alpha,1}(t)$ (Fig. \ref{fig:sinvne1}(b)). 
Within the specific time window shown in Fig. \ref{fig:sinvne1}, we observe irregular dynamics for $\alpha = 1.0$. However, for {$\alpha =\beta= 0.5$ or $0.4$ (finely dotted red and solid blue lines, respectively)}, the fractional orders slow down the dynamics, and the onset of non-periodicity takes longer to appear.
The combination $(\alpha,\beta)=(0.4,1.0)$  or $(\alpha,\beta)=(0.5,1.0)$ exhibits a {more regular} dynamics after a transient time.
This suggests that the fractional order can stabilize otherwise non-periodic evolution.}

\begin{figure}[t]
\centering
\includegraphics[scale=0.5]{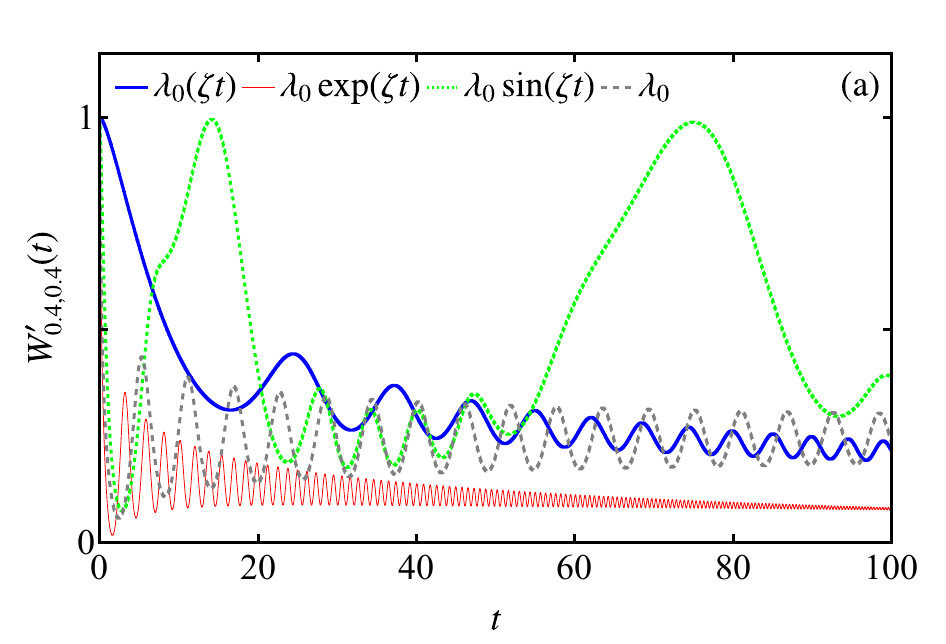}
\hspace{0.5cm}
\includegraphics[scale=0.5]{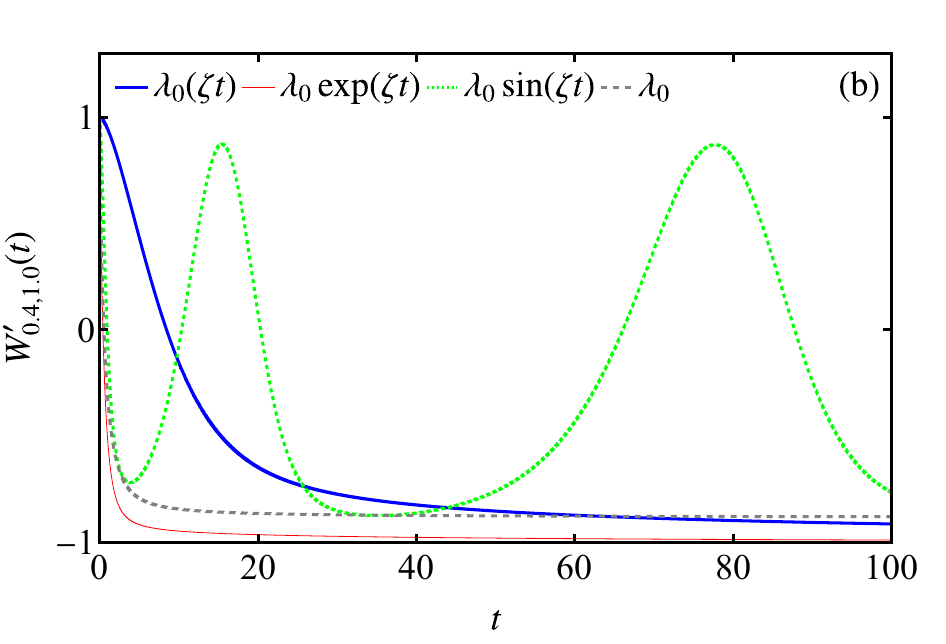}
  \caption{
The normalized population inversion as a function of $t$, setting $a_0=1$, $b_0=0$, $n=0$, $\lambda_0=1$, $p=1.0$, $\Delta=0.5$, $\alpha=0.4$, $\zeta=0.16$ (for the linear and exponential modulations) and $\zeta=1.0$ (for the sinusoidal modulation).
The coupling parameters considered are:  
$\lambda(t)=\lambda_0 (\zeta t)$ (thick solid blue line), $\lambda(t)=\lambda_0 e^{\zeta t}$ (thin solid red line) and $\lambda(t)=\lambda_0 \sin(p \zeta t)$ (dotted green line), $\lambda(t)=\lambda_0$ (dashed gray line). 
Panel (a) is for $\beta=\alpha$, and panel (b) is for $\beta=1.0$.
  } 
  \label{fig:comparison}
\end{figure}

To offer a final perspective on the comparison of the modulations of the coupling parameter and to encompass a longer time interval, we show the different modulations for fixed $\alpha=0.4$ in Fig. 
\ref{fig:comparison}. 
When $\alpha=\beta$ (Fig. \ref{fig:comparison}(a)),  for the constant {(dashed gray line)}, linear {(thick solid blue line)}, and exponential modulations {(thin solid blue line)}, we observe a somewhat similar behavior but with a different time scale and amplitude. 
In particular, within the considered time interval, the exponential modulation causes a population inversion closer to $0$, with an almost steady value {reached in} latter times.
Furthermore, the non-periodicity that results from the sinusoidal coupling {(dotted green line)} stands out.
When $\alpha=\beta$ (Fig. \ref{fig:comparison}(b)), the decay is exhibited for all cases, except, again, the sinusoidal one.
This is remarkable, as the sinusoidal coupling contradict{s} the behavior observed in the other curves, in which the time dependence appears to influence only the timescale of the dynamics.

\section{Conclusions}
\label{sec:conc}

We have presented a comprehensive study of the TDJC model within the framework of fractional quantum mechanics, by incorporating Caputo fractional time derivatives and exploring two distinct formulations for the imaginary unit in the TFSE: one involving $i^\alpha$ and another preserving the canonical $i$.
This approach allows the investigation of memory effects in the coherent dynamics of light–matter interaction.

Our results demonstrate that fractional time evolution induces qualitative modifications to the system’s dynamics, such as non-conservation of probability, damping of {RO}, and long-time asymptotic behavior. 
For the exponent of the imaginary unit equal to the fractional order ($\beta = \alpha < 1$), we observe damped oscillations that persist around a stationary value. 
On the other hand, when the imaginary unit remains unchanged ($\beta = 1$), the system experiences decay in population inversion and entanglement, resembling the dissipative processes found in open quantum systems.
These effects are inherited by the population inversion and VNE, 
resulting in quantities that oscillate around a fixed value ($\alpha=\beta\neq1$) 
or decay to an asymptotic {one} ($\alpha\neq\beta$ and $\alpha<1$).
Additionally, we investigated how the parameters of the system can influence the behavior of the {normalization parameter $\mathcal{P}_{\alpha,\beta}(t)$},  {finding that detuning influences its behavior, especially at weaker coupling}.

Considering a linear ramp in the modulation, 
we observed that RO initiates slowly but presents an acceleration over time. 
Remarkably, when {$\alpha =0.7$ and $ \beta = 1$},  we observe, in some instances, values closer to the minimum in both the population inversion and the VNE than those obtained in the $\alpha = \beta = 1$ scenario.
As a consequence of the exponential coupling parameter, the RO are considerably denser than in the constant coupling scenario, 
accompanied by a subtle change in the amplitude.
{
This asymptotic entanglement value differs from the ones found for constant or linear coupling,} as the dynamics cease at a different point.
In particular, sinusoidal coupling, which models atoms traversing standing-wave cavities, leads to non-periodic dynamics that become regularized when the fractional order is decreased, highlighting a novel  {FC-induced} stabilization mechanism.  
From a physical perspective, the fractional formalism offers an alternative approach to investigating { environment influence} without resorting to explicit bath models.

{Regarding research gaps, we address two points: first, this work focuses specifically on the $\beta=1$ and $\beta=\alpha$ regimes, which correspond to the two established formulations of the TFSE \cite{Naber2004}. 
A direct extension would be to explore dynamics in broader regions of the pair $(\alpha,\beta)$ and to investigate how these parameters influence the non-periodicity of the solutions.
Second, we did not perform quantitative simulations of non-Markovian effects; consequently, mapping the fractional parameters to memory parameters in other non-Markovian frameworks remained an open question.}

The proposed time fractional TDJC model, despite its theoretical character, may inspire experimental realizations. For instance, in cavity QED, the sinusoidal coupling $\lambda(t) = \lambda_0 \sin(p\zeta t)$ capture{s} the configuration of atoms traversing cavity standing waves. 
Deviations from periodic behavior due to memory effects can be investigated through RO measurements. 
In superconducting circuit QED, transmon-cavity systems facilitate the dynamic modulation of parameters such as $\lambda(t)$, $\Delta$, and $\omega$. 
This creates a possibility of exploring non-monotonic decay and memory-modified entanglement dynamics. 
Trapped-ion quantum simulators, through stroboscopic control of JC-like Hamiltonians, can emulate fractional evolution, thereby facilitating the verification of transitions between periodic and aperiodic regimes. Furthermore, dissipative solid-state systems, including quantum dots and plasmonic cavities, provide a natural setting for testing TFSE-based models by analyzing sub-exponential relaxation data.

Finally, we hope that our discussion and results, encompassing the off-resonance regime and multiple time-dependent couplings, will further advance the practical applications of FC.

\section*{Acknowledgments}
The authors thank Dr. Alison A. Silva for helpful discussions. The authors thank the financial support from the Brazilian Federal Agencies (CNPq), grants 407299/2018-1, 311168/2020-5, and 313124/2023-0; the S\~ao Paulo Research Foundation (FAPESP) under grants 2024/05700-5; 
Coordenação de Aperfeiçoamento de Pessoal de Nível Superior (CAPES, Finance Code 001), and 
Funda\-\c c\~ao A\-rauc\'aria, Project No. 305. D. C. acknowledges financial support from Instituto Serrapilheira, and the Pró-Reitoria de Pesquisa e Inovação (PRPI) from the Universidade de São Paulo (USP) by financial support through the Programa de Estímulo à Supervisão de Pós-Doutorandos por Jovens Pesquisadores. E.C.G acknowledges the financial support from FAPESP under grant 2025/02318-5.
\section*{CRediT authorship contribution statement}
{\bf Enrique Gabrick}: Conceptualization, Methodology, Software, Validation, Formal analysis, Investigation, Writing - Original Draft, Writing - Review \& Editing, Funding acquisition.
{\bf Thiago Tsutsui}: Conceptualization, Methodology, Software, Validation, Formal analysis, Investigation, Writing - Original Draft, Writing - Review \& Editing, Funding acquisition.
{\bf Danilo Cius}: Conceptualization, Methodology, Validation, Investigation, Writing - Original Draft, Writing - Review \& Editing, Funding acquisition.
{\bf Ervin Lenzi}: Conceptualization, Methodology, Validation, Formal analysis, Writing - Review \& Editing, Funding acquisition, Supervision, Project administration.
{\bf Antonio de Castro}: Conceptualization, Methodology, Writing - Review \& Editing, Funding acquisition, Supervision, Project administration.
{\bf Fabiano Andrade}: Conceptualization, Methodology, Writing - Review \& Editing, Funding acquisition, Supervision, Project administration.
\section*{Data availability}
The data supporting the findings of this study are available within the article. 
\section*{Declaration of competing interest}
The authors declare that they have no known competing financial interests or personal relationships that could have appeared to influence the work reported in this paper.
\bibliographystyle{elsarticle-num}
\bibliography{Ref_with_fixed_titles}

\end{document}